\documentclass[amsmath,amssymb,amsbsy,prb,reprint,superscriptaddress]{revtex4-1}

\usepackage{graphicx}
\usepackage{epsfig}
\usepackage{rotating}
\usepackage{dcolumn}
\usepackage{bm}
\usepackage{mathrsfs}
\usepackage{leftidx}

\begin{document}

\title{Influence of Dzyaloshinskii-Moriya interactions on magnetic structure of a spin-1/2 deformed kagome lattice antiferromagnet}

\author{Kyusung Hwang}
\affiliation{Department of Physics, University of Toronto, Toronto, Ontario M5S 1A7, Canada}

\author{Kwon Park}
\affiliation{School of Physics, Korea Institute for Advanced Study, Seoul 130-722, Korea}

\author{Yong Baek Kim}
\affiliation{Department of Physics, University of Toronto, Toronto, Ontario M5S 1A7, Canada}
\affiliation{School of Physics, Korea Institute for Advanced Study, Seoul 130-722, Korea}


\date{\today}

\begin{abstract}
Motivated by the recent neutron scattering experiment on Rb$_2$Cu$_3$SnF$_{12}$ [Nat.~Phys.~${\bf 6}$, 865 (2010)], 
we investigate the effect of Dzyaloshinskii-Moriya interactions in a theoretical model for the magnetic structure of this material.
Considering the valence bond solid ground state, which has a 12-site unit cell, we develop the bond operator mean-field theory.
It is shown that the Dzyaloshinskii-Moriya interactions significantly modify the triplon dispersions around the $\Gamma$ point and cause a shift of the spin gap (the minimum triplon gap) position from the K to $\Gamma$ point in the first Brillouin zone.
The spin gap is also evaluated in exact diagonalization studies on a 24-site cluster.
We discuss a magnetic transition induced by the Dzyaloshinskii-Moriya interactions in the bond operator framework.
Moreover, the magnetization process under external magnetic fields is studied within the exact diagonalization and strong coupling expansion approaches.
We find that the results of all above approaches are consistent with the experimental findings.
\end{abstract}

\maketitle

\section{INTRODUCTION\label{sec:INTRODUCTION}}
Frustrated quantum magnets have been studied intensively to unveil possible novel phenomena which can be caused by the interplay of frustration and quantum effects. 
Spin-1/2 two dimensional (2D) antiferromagnet on the kagome lattice is a central example of such frustrated quantum magnets. 
On the theoretical front, various studies have suggested several candidate ground states for the magnetically disordered phase of the spin-1/2 antiferromagnetic Heisenberg model: 
a variety of spin liquids
\cite{book_Misguich_Lhuillier,
SL_Sachdev_1992,
SL_Hastings_2000,
SL_Wang_2006,
SL_Ran_2007,
SL_Ryu_2007,
SL_Hemele_2008,
DMRG_Jiang_2008,
DMRG_Yan_2011,
PWF_Iqbal_2011}
and valence bond solids
\cite{VBS_Marston_1991,
VBS_Nikolic_2003,
VBS_Singh_2007,
VBS_Singh_2008,
VBS_Yang_2008,
MERA_Evenbly_2010,
VMC_Iqbal_2011,
VBS_Hwang_2011}.
Experimentally, a lot of effort has been made to realize the material system of such geometrically frustrated antiferromagnet. 
The Herbertsmithite
\cite{herbertsmithite_Shores_2005,
herbertsmithite_Ofer_2006,
herbertsmithite_Helton_2007,
herbertsmithite_Mendels_2007,
impurity_NMR_Olariu_2008,
DM_ESR_Zorko_2008,
herbertsmithite_Helton_2010,
herbertsmithite_Han_2012}
ZnCu$_3$(OH)$_6$Cl$_2$ has the most ideal structure with uniform exchange couplings and shows no magnetic ordering down to much lower temperatures than the Curie-Weiss temperature.
On the other hand, there may be several additional factors that would affect the nature of the ground state.
For example, the Dzyaloshinskii-Moriya interaction and impurity spins may be present in the material.
\cite{DM_impurity_Rigol_2007,
impurity_NMR_Olariu_2008,
DM_ESR_Zorko_2008,
DM_impurity_Rousochatzakis_2009,
DM_NMR_Jeong_2011}
Because of these complications, it has been difficult to pin down the true ground state of the system.
These considerations also suggest that it is important to understand the effect of various perturbations to the
ideal kagome lattice structure and magnetic anisotropies to identify the true ground state of the system.

\begin{figure}
 \centering
 \includegraphics[width=0.9\linewidth]{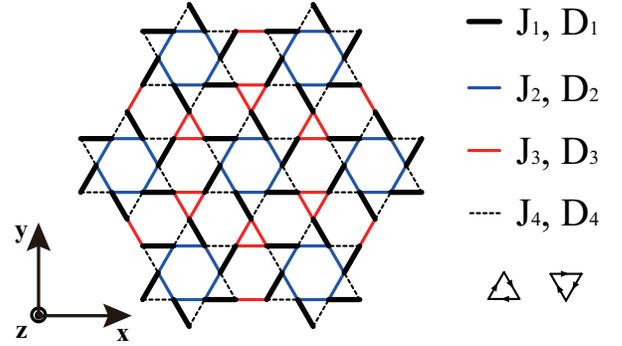}
 \caption{(Color online) The deformed kagome lattice with four different couplings for the Heisenberg and Dzyaloshinskii-Moriya interactions respectively.
 The DM vectors are assumed to be perpendicular to the lattice plane
 and their direction is along the positive direction of the z-axis 
 when the orientation from $i$ to $j$ in $\mathrm{\bf S}_i  \times \mathrm{\bf S}_j$ is clockwise as indicated by the arrows in the lower right triangles.
 For the VBS state with the 12-site unit cell, the valence bonds are formed in the interaction links with the strongest couplings, $J_1$ and $D_1$ (thick black).
 \label{fig:deformed_kagome_lattice}}
\end{figure}

Rb$_2$Cu$_3$SnF$_{12}$ is another compound that was synthesized as an attempt to materialize the kagome lattice antiferromagnet.
Unlike the Herbertsmithite, however, it has a deformed kagome lattice structure (see Fig. \ref{fig:deformed_kagome_lattice}). 
Magnetic susceptibility and magnetization measurements show that it is a magnetically-disordered 2D spin-gapped system.\cite{suscep_magnetization_gap}
As a model to describe the system, a spin-1/2 antiferrmagnetic Heisenberg model was considered with four different exchange couplings due to its deformed structure, labeled by $J_1$, $J_2$, $J_3$, and $J_4$. 
The exchange coupling constants are estimated by fitting the thermodynamic data: $J_1$, $J_2$, and $J_3$ are quite similar in magnitude ($\sim$200 K), while $J_4$ is about half of the others ($J_1 > J_2 > J_3 \gg J_4$).
The size of the spin gap is estimated to be $\Delta/k_B$=21 K.

The recent neutron scattering experiment in conjunction with a dimer series expansion study has shown that the ground state of Rb$_2$Cu$_3$SnF$_{12}$ is indeed the valence bond solid (VBS) with a 12-site unit cell.\cite{neutron_series}
The valence bonds are formed in the strongest interaction links with the coupling constant $J_1$ (black line segments in Fig. \ref{fig:deformed_kagome_lattice}). 
Further, it turned out that the Dzyaloshinskii-Moriya (DM) interactions are not negligible in Rb$_2$Cu$_3$SnF$_{12}$ as seen in the triplon dispersions.
If there are only (isotropic) Heisenberg interactions, the system has SO(3) global spin rotation symmetry so that its triplon dispersions should be triply-degenerate.
When there are (anisotropic) DM interactions, the degeneracy of triplon dispersions is lifted.
It was observed in neutron scattering experiment that the triplon dispersions of Rb$_2$Cu$_3$SnF$_{12}$ are only doubly-degenerate.
This fact indicates that there exist significant DM interactions in Rb$_2$Cu$_3$SnF$_{12}$.

The valence bond solid state with a 12-site unit cell was studied by some of us for the spin-1/2 antiferromagnetic Heisenberg model on the deformed kagome lattice.\cite{Yang_BOMFT}
In that study, the weakest interaction was controlled with $J_4=\alpha J$ and $J_1=J_2=J_3=J$.
The VBS state with a 12-site unit cell turned out to have lower energy than the VBS state with a 36-site unit cell, 
which was suggested for uniform kagome lattice antiferromagnet ($\alpha=1$ case), 
when $\alpha$ is less than 0.97 within the bond operator mean-field theory.
Concerning Rb$_2$Cu$_3$SnF$_{12}$, the triplon dispersions were predicted as well.
The predicted triplon energy bands are triply-degenerate due to the SO(3) symmetry of the Heisenberg Hamiltonian and spin gap is found at the K point.
However, the spin gap (the minimum triplon gap) was observed at $\Gamma$ point in the recent neutron scattering experiment.\cite{neutron_series}
This presents another evidence for the importance of the DM interaction in Rb$_2$Cu$_3$SnF$_{12}$.

In this paper, we theoretically investigate the effect of the Dzyaloshinskii-Moriya interaction in the deformed kagome lattice antiferromanget via various methods.
The model Hamiltonian for the system consists of the Heigenberg ($\mathcal{H}_{J}$) and Dzyaloshinskii-Moriya ($\mathcal{H}_{D}$) interactions between the nearest neighbors;
\begin{eqnarray}
   \mathcal{H}_{J} + \mathcal{H}_{D}
   = \sum_{\left< i,j \right>} J_{ij} \mathrm{\bf S}_i \cdot \mathrm{\bf S}_j
   + \sum_{\left< i,j \right>} \mathrm{\bf D}_{ij} \cdot \mathrm{\bf S}_i  \times \mathrm{\bf S}_j.
 \label{eq:Hamiltonian}
\end{eqnarray}
For the Heisenberg interactions, we consider four different exchange couplings, $J_1 > J_2 > J_3 \gg J_4$, depending on the interaction link as shown in Fig. \ref{fig:deformed_kagome_lattice}.
As for the DM interactions, we use the result from series expansion study.\cite{neutron_series}
According to the series expansion, the DM vectors $\mathrm{\bf D}_{ij}$ have small in-plane components compared to the out-of-plane component, and the overall behavior of excitation spectrum is determined by the out-of-plane component whereas the in-plane ones make minor corrections.
Due to this reason, the DM vectors are assumed to have only the out-of-plane component in our study ($\mathrm{\bf D}_{ij} // \hat{z}$).
Like the Heisenberg interactions, we consider four different DM interactions, the magnitude of which is labeled by $D_n ~ (n=1,\cdots, 4)$.
$D_n$ is assumed to be proportional to $J_n$, \emph{i.e.} $D_n/J_n = d_z ~ (n=1,\cdots,4)$.
The strength of the DM interactions is controlled by the parameter $d_z$ in this paper.
The direction of $\mathrm{\bf D}_{ij}$ is determined by the Moriya's rule\cite{DM_int,DM_kagome,DM_ED} to be along the positive z-direction if the orientation from the site $i$ to $j$ in $\mathrm{\bf S}_i  \times \mathrm{\bf S}_j$ is clockwise in each triangle of the kagome lattice.
In the lower right part of Fig. \ref{fig:deformed_kagome_lattice}, the orientations from $i$ to $j$ for the positive DM vector are denoted with arrows in two possible types of triangles in the kagome lattice.
It must be noted that the model Hamiltonian has SO(2) global spin rotation symmetry in the xy-plane because every DM vector is along the z-direction.
Therefore, the generator of spin rotation in the xy-plane, $\sum_i {S}_{i,z}$, is a good quantum number of the Hamiltonian.

To analyze this Hamiltonian, we take three different approaches: bond operator mean-field theory, exact diagonalization, and strong coupling expansion.
First, the bond operator mean-field theory is used to investigate the effect of the DM interaction on the triplon dispersions.
Within the bond operator theory, we consider the VBS state with the 12-site unit cell in Fig. \ref{fig:dimer_configuration} as the ground state and obtain the triplon dispersions. These disperions are shown to be consistent with the results of the neutron scattering experiment.
The dispersions are twofold degenerate, which results from the SO(2) symmetry of the Hamiltonian.
By increasing the strength of the DM interactions ($d_z$) from zero, 
we observe that the position of the spin gap (the minimum triplon gap) in the first Brillouin zone shifts from the K to $\Gamma$ point at $d_z=0.09$.
Beyond $d_z=0.09$, the size of the spin gap decreases gradually, but the position of spin gap does not change from the K point.
It is interesting to note that the spin gap is linearly decreasing as a function of $d_z$ after $d_z=0.09$.
By performing linear fitting, we estimate the critical value $d_{z,c}=0.381$, at which the spin gap becomes zero.
The value of $d_z$ in Rb$_2$Cu$_3$SnF$_{12}$ was estimated to be 0.18 in the series expansion study.
At $d_z=0.18$, the spin gap in the bond operator theory is 0.464 $J_1$.
This value is about four times larger than the neutron scattering result, 0.126 $J_1$.
But, it is not surprising because the bond operator mean-filed theory tends to overestimate the spin gap. 

Secondly, we conduct the exact diagonalization study for finite size clusters in order to calculate the spin gap more accurately.
In the 24-site cluster, the spin gap is found to be 0.229 $J_1$, which is not far from the experimental value.
As a matter of fact, this value is the best among the results of various approaches in our study.
Via the exact diagonalization, we also study magnetization process because magnetically-disordered spin-gapped systems often show interesting behaviors in magnetization process.
Exact diagonalization is restricted to small size clusters due to the large unit cell of the deformed kagome lattice.
To overcome this limitation, we also employ strong coupling expansion approach.
Strong coupling expansion is a degenerate perturbation theory which provides an effective Hamiltonian near the magnetic fields strong enough to close the spin gap.
This approach allows the larger system sizes than the clusters used in the exact diagonalization study because the full Hilbert space is projected into the subspace of the states relevant to the magnetization process.
From the exact diagonalization and strong coupling expansion studies, we obtain qualitatively the same magnetization curves.
The magnetization curves increase continuously from zero to saturated magnetization as the magnetic field is increased, which is consistent with the magnetization measurement.

The outline of the rest of the paper is as follows.
In Sec. \ref{sec:BOMFT}, we develop the bond operator mean-field theory for the deformed kagome lattice antiferromanget.
Exact diagonalization study is presented in Sec. \ref{sec:ED}.
In Sec. \ref{sec:SCE}, we perform the strong coupling expansion.
Lastly, results are compared with the recent experimental results and summarized in Sec. \ref{sec:DISCUSSION}.

\section{BOND OPERATOR MEAN-FIELD THEORY\label{sec:BOMFT}}
\begin{figure}
 \centering
 \includegraphics[width=0.7\linewidth]{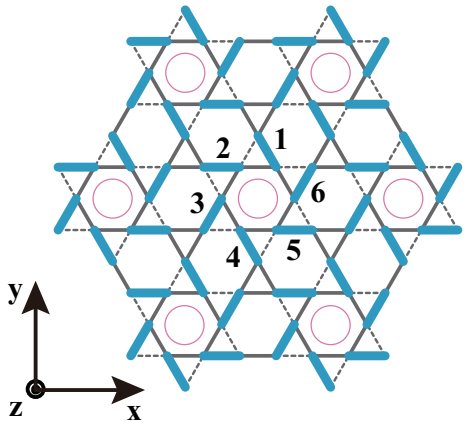}
 \caption{(Color online) The dimer configuration for the VBS state with the 12-site unit cell.
 Dimers are denoted with the blue thick line segments.
 The 12-site unit cell contains six dimers labeled by $1,\cdots,6$.
 In our convention, ${\bf S}_R$ in a dimer is located at the vertices of the hexagon denoted with circle.
 \label{fig:dimer_configuration}}
\end{figure}

In this section, we construct the bond operator mean-field theory for the VBS state with the 12-site unit cell.
The dimer configuration for the VBS state is depicted in Fig. \ref{fig:dimer_configuration}, 
where dimers are denoted with thick blue line segments and six dimers in the 12-site unit cell are labeled by $1,\cdots,6$.

\subsection{Bond operator representation}

To introduce the basis for the bond operator representation, we consider the following Hamiltonian of a pair of spins, ${\bf S}_L$ and ${\bf S}_R$, which corresponds to an isolated spin dimer with the coupling constants $J_1$ and $D_1$ in Fig. \ref{fig:dimer_configuration};
\begin{eqnarray}
  \mathcal{H}_{dimer} 
  &=& J_1 {\bf S}_L \cdot {\bf S}_R + D_1 \hat{z} \cdot {\bf S}_L \times {\bf S}_R,
  \nonumber\\
  &=& \frac{1}{2} (J_1+iD_1) {S}_{L,+} {S}_{R,-} + \frac{1}{2} (J_1-iD_1) {S}_{L,-} {S}_{R,+} 
  \nonumber\\  
  &+& J_1 {S}_{L,0} {S}_{R,0},
  \label{H_dimer}
\end{eqnarray}
where ${S}_{n,\pm}={S}_{n,x} \pm i {S}_{n,y}$ and ${S}_{n,0}={S}_{n,z}$ $(n=L,R)$.
The energy eigenstates are given by
\begin{subequations}
\label{eq:BOT_basis}
\begin{eqnarray}
 s^{\dagger} \left| 0 \right> &=& \left| s \right> = \frac{1}{\sqrt{2}} ( e^{i \alpha/2} \left| \uparrow \downarrow \right> - e^{-i \alpha/2} \left| \downarrow \uparrow \right> ),
 \\
 t_{+}^{\dagger} \left| 0 \right> &=& \left| t_{+} \right> = - \left| \uparrow \uparrow \right>,
 \\
 t_{0}^{\dagger} \left| 0 \right> &=& \left| t_{0} \right> = \frac{1}{\sqrt{2}} ( e^{i \alpha/2} \left| \uparrow \downarrow \right> + e^{-i \alpha/2} \left| \downarrow \uparrow \right> ),
 \\
 t_{-}^{\dagger} \left| 0 \right> &=& \left| t_{-} \right> = \left| \downarrow \downarrow \right>,
\end{eqnarray}
\end{subequations}
where 
\begin{eqnarray}
 e^{i \alpha} = \frac{J_1+iD_1}{\sqrt{J_1^2+D_1^2}}.
\end{eqnarray}
The corresponding energies are
\begin{subequations}
\begin{eqnarray}
 \epsilon_s &=& -\frac{1}{4}J_1-\frac{1}{2}\sqrt{J_1^2+D_1^2},
 \\
 \epsilon_{t_{0}} &=& -\frac{1}{4}J_1+\frac{1}{2}\sqrt{J_1^2+D_1^2},
 \\
 \epsilon_{t_{\pm}} &=& \frac{1}{4}J_1.
\end{eqnarray}
\label{eq:H_dimer_energies}
\end{subequations}
It must be noted that $s^{\dagger} \left| 0 \right>$ is not the pure spin singlet state and $t_{0}^{\dagger} \left| 0 \right>$ is not one of the pure spin triplet states unless $\alpha = 0$ (or $D_1=0$).
$s^{\dagger}$, $t_{+}^{\dagger}$, $t_{0}^{\dagger}$, and $t_{-}^{\dagger}$ are operators to create a bond particle in the associated state and satisfy the bosonic commutation rules.
The subscript of $t$-operator, $\left\{ +, 0, - \right\}$, implies the quantum number of $S_{L,z}+S_{R,z}$, which originates from the SO(2) spin rotation symmetry of the Hamiltonian in Eq. (\ref{H_dimer}).
In other words, $\left\{ +, 0, - \right\}$ indicates the magnetic quantum number of the $t$-bosons.

With the above bond operators, the spin operators, ${\bf S}_L$ and ${\bf S}_R$, can be rewritten as follows;
\begin{subequations}
\label{eq:bond_op}
\begin{eqnarray}
 {S}_{L,+} &=& \frac{1}{\sqrt{2}} ( s^{\dagger} t_{-} + t_{+}^{\dagger} s - t_{+}^{\dagger} t_{0} + t_{0}^{\dagger} t_{-} ) \cdot e^{-i \alpha/2},
 \\
 {S}_{L,-} &=& \frac{1}{\sqrt{2}} ( s^{\dagger} t_{+} + t_{-}^{\dagger} s - t_{0}^{\dagger} t_{+} + t_{-}^{\dagger} t_{0} ) \cdot e^{i \alpha/2},
 \\
 {S}_{L,0} &=& \frac{1}{2} ( s^{\dagger} t_{0} + t_{0}^{\dagger} s + t_{+}^{\dagger} t_{+} - t_{-}^{\dagger} t_{-} ),
 \\
 {S}_{R,+} &=& \frac{1}{\sqrt{2}} ( - s^{\dagger} t_{-} - t_{+}^{\dagger} s - t_{+}^{\dagger} t_{0} + t_{0}^{\dagger} t_{-} ) \cdot e^{i \alpha/2},
 \\
 {S}_{R,-} &=& \frac{1}{\sqrt{2}} ( - s^{\dagger} t_{+} - t_{-}^{\dagger} s - t_{0}^{\dagger} t_{+} + t_{-}^{\dagger} t_{0} ) \cdot e^{-i \alpha/2},
 \\
 {S}_{R,0} &=& \frac{1}{2} ( - s^{\dagger} t_{0} - t_{0}^{\dagger} s + t_{+}^{\dagger} t_{+} - t_{-}^{\dagger} t_{-} ).
\end{eqnarray}
\end{subequations}
In the above representation, we set $\hbar=1$.
There are two reasons why we choose the states (\ref{eq:BOT_basis}) as the basis for the bond operator representation against the conventional one.\cite{Sachdev_Bhatt,spin_ladder_Gopalan}
First, if we use the conventional basis, 
the DM interactions with the coupling constant $D_1$ do not appear in the quadratic part of the Hamiltonian.
By taking the above basis, those DM interactions can be included from the beginning.
Secondly, the ground state cannot be a pure spin singlet state due to the broken spin rotation symmetry in the presence of the DM interactions.
In the conventional representation, or when $\alpha=0$, $t$-bosons are sometimes called the triplons because they represents the spin triplet states.
Although all of the $t$-bosons in our representation are not in the pure spin triplet states, we will call them the triplons throughout the paper.

\subsection{Mean-field Hamiltonian}
In order to write the Hamiltonian (\ref{eq:Hamiltonian}) in terms of the bond operators,
it is helpful to arrange the Hamiltonian into the following form;
\begin{eqnarray}
  \mathcal{H}_{J}+\mathcal{H}_{D} = \sum_{n=1}^{4} \mathcal{H}_n,
 \label{eq:H_J_H_DM} 
\end{eqnarray}
where $\mathcal{H}_n$ includes the interactions with coupling constants $J_n$ and $D_n$.
That is,
\begin{subequations}
\label{eq:H_n}
\begin{eqnarray}
  \mathcal{H}_1
  =
  \sum_{\bf r} \sum_{m=1}^{6}
  &&
  \left[ J_1 {\bf S}_{L}^m ({\bf r}) \cdot {\bf S}_{R}^{m} ({\bf r}) \right.
  \nonumber\\  
  &+& \left. D_1 \hat{z} \cdot {\bf S}_{L}^m ({\bf r}) \times {\bf S}_{R}^{m} ({\bf r}) \right],
\end{eqnarray}
\begin{eqnarray}
  \mathcal{H}_2
  =
  \sum_{\bf r} \sum_{m=1}^{6}
  &&
  \left[ J_2 {\bf S}_{R}^m ({\bf r}) \cdot {\bf S}_{R}^{m+1} ({\bf r}) \right.
  \nonumber\\  
  &+& \left. D_2 \hat{z} \cdot {\bf S}_{R}^m ({\bf r}) \times {\bf S}_{R}^{m+1} ({\bf r}) \right],
\end{eqnarray}
\begin{eqnarray}
  \mathcal{H}_3
  =
  \sum_{\bf r} \sum_{(m,n;{\bf R}) \in I_3}
  &&
  \left[ J_3 {\bf S}_{L}^m ({\bf r}) \cdot {\bf S}_{L}^{n} ({\bf r}+{\bf R}) \right.
  \nonumber\\  
  &+& \left. D_3 \hat{z} \cdot {\bf S}_{L}^m ({\bf r}) \times {\bf S}_{L}^{n} ({\bf r}+{\bf R}) \right],
\end{eqnarray}
\begin{eqnarray}
  \mathcal{H}_4
  =
  \sum_{\bf r} \sum_{m=1}^{6}
  &&
  \left[ J_4 {\bf S}_{R}^{m+1} ({\bf r}) \cdot {\bf S}_{L}^{m} ({\bf r}) \right.
  \nonumber\\  
  &+& \left. D_4 \hat{z} \cdot {\bf S}_{R}^{m+1} ({\bf r}) \times {\bf S}_{L}^{m} ({\bf r}) \right],
\end{eqnarray}
\end{subequations}
where ${\bf S}_{R}^{7}={\bf S}_{R}^{1}$, and the set $I_3$ is given by
\begin{eqnarray}
  I_3=
  &&
  \left\{
  (1,5;{\bf r}_C),
  (3,1;-{\bf r}_B),
  (5,3;{\bf r}_A),
  \right.
  \nonumber\\
  &&
  \left.
  (2,6;-{\bf r}_A),
  (4,2;-{\bf r}_C),
  (6,4;{\bf r}_B)
  \right\}.
  \label{eq:I_3}
\end{eqnarray}
In Eq. (\ref{eq:H_n}) and (\ref{eq:I_3}), ${\bf r}$ implies the lattice vector, $m$ and $n$ are dimer indices within the unit cell, $L$ and $R$ are indices to distinguish spins in a given dimer, and
\begin{eqnarray}
  &&
  {\bf r}_A = 4 a \hat{x},
  \nonumber\\
  &&
  {\bf r}_B = 4 a \left( \frac{1}{2} \hat{x} + \frac{\sqrt{3}}{2} \hat{y} \right),
  \nonumber\\
  &&
  {\bf r}_C = {\bf r}_B - {\bf r}_A,
\end{eqnarray} 
where $a$ is the lattice spacing.
In our convention, spin $R$ is located at the vertices of the hexagon denoted with circle in Fig. \ref{fig:dimer_configuration}.
Rewriting the Hamiltonian (\ref{eq:H_n}) in terms of the bond operators is straightforward.
In addition, the following hardcore constraint is required at each dimer to restrict the Hilbert space into the physical one;
\begin{eqnarray}
 s^{\dagger} s + \sum_{l=+,0,-} t_{l}^{\dagger} t_{l} = 1.
 \label{eq:hardcore_constraint}
\end{eqnarray}
Moreover, the operators $s$ and $s^{\dagger}$ are replaced by a number $\bar{s}$ to describe the VBS order.
$\bar{s}$ denotes the condensate density of the $s$-bosons.
Therefore, we consider the following Hamiltonian in the bond operator theory;
\begin{eqnarray}
 \mathcal{H} 
 &=& \mathcal{H}_{J} + \mathcal{H}_{D}
 \nonumber\\
 &-& \sum_{\bf r} \sum_{m=1}^{6} \mu \left[ \bar{s}^2 + \sum_{l=+,0,-} t_{l,m}^{\dagger}({\bf r}) t_{l,m}({\bf r}) - 1 \right],
 \label{eq:BOT_Hamiltonian}
\end{eqnarray}
where ${\bf r}$ is the lattice vector, $m$ dimer index within the unit cell, and $\mu$ the Lagrange multiplier for the hardcore constraint.
In principle, the condensate density $\bar{s}$ and Lagrange multiplier $\mu$ can depend on ${\bf r}$ and $m$.
However, based on the symmetry of the underlying lattice, they are taken to be uniform on the lattice.
Arranging the Hamiltonian with respect to the degree of $t$-boson operators,
\begin{eqnarray}
 \mathcal{H}=N_{uc} \epsilon_o + \mathcal{H}_{quad} + \mathcal{H}_{quartic},
\end{eqnarray}
where $N_{uc}$ is the number of unit cells,
\begin{eqnarray}
 \epsilon_o = 6 \left[ \epsilon_{s} \bar{s}^2 + \mu (1-\bar{s}^2) \right],
\end{eqnarray}
and $\mathcal{H}_{quad}$ and $\mathcal{H}_{quartic}$ consist of quadratic and quartics terms of $t$-operators, respectively.
Cubic terms are ignored in the above Hamiltonian because magnetically disordered VBS states are of interest here.

The quadratic part of the Hamiltonian is considered first.
After Fourier transformations, it can be written as follows.
\begin{eqnarray}
  \mathcal{H}_{quad}=\mathcal{H}_{quad,1}+\mathcal{H}_{quad,2}+\mathcal{H}_{quad,3}+\mathcal{H}_{quad,4},
\end{eqnarray}
where
\begin{subequations}
\label{eq:H_quad_n}
\begin{eqnarray}
  \mathcal{H}_{quad,1}&=&\sum_{l=+,0,-}\sum_{\bf k} \sum_{m=1}^{6} (\epsilon_{t_l}-\mu) t_{l,m}^{\dagger} ({\bf k}) t_{l,m} ({\bf k}),
  \nonumber\\
\end{eqnarray}
\begin{eqnarray}
 &&\mathcal{H}_{quad,2}
 \nonumber\\
 &=&
 \sum_{l=+,0,-} \sum_{\bf k} \sum_{m=1}^{6}
 \frac{1}{4} (J_2+l \cdot iD_2) {\bar s}^2 t_{l,m}^{\dagger}({\bf k}) t_{l,m+1}({\bf k})
 \nonumber\\
 &+& h.c.
 \nonumber\\
 &+&
 \sum_{l=+,0,-} \sum_{\bf k} \sum_{m=1}^{6}
 \frac{1}{4} (J_2+l \cdot iD_2) {\bar s}^2 t_{l,m}^{\dagger}({\bf k}) t_{-l,m+1}^{\dagger}(-{\bf k})
 \nonumber\\
 &+& h.c.,
\end{eqnarray}
\begin{eqnarray}
 && \mathcal{H}_{quad,3}
 \nonumber\\
 &=&
 \sum_{l=+,0,-} \sum_{\bf k} \sum_{(m,n;{\bf R}) \in I_3}
 \frac{1}{4} (J_3+l \cdot iD_3) {\bar s}^2 e^{i{\bf k} \cdot {\bf R}}
 \nonumber\\
 &&~~~~~~~~~~~~~~~~~~~~~~~~~~~
 \cdot
 t_{l,m}^{\dagger}({\bf k}) t_{l,n}({\bf k})
 \nonumber\\
 &+& h.c.
 \nonumber\\
 &+&
 \sum_{l=+,0,-} \sum_{\bf k} \sum_{(m,n;{\bf R}) \in I_3}
 \frac{1}{4} (J_3+l \cdot iD_3) {\bar s}^2 e^{i{\bf k} \cdot {\bf R}} 
 \nonumber\\
 &&~~~~~~~~~~~~~~~~~~~~~~~~~~~~
 \cdot 
 t_{l,m}^{\dagger}({\bf k}) t_{-l,n}^{\dagger}(-{\bf k})
 \nonumber\\
 &+& h.c.,
\end{eqnarray}
\begin{eqnarray}
 && \mathcal{H}_{quad,4}
 \nonumber\\
 =
 &-&
 \sum_{l=+,0,-} \sum_{\bf k} \sum_{m=1}^{6}
 \frac{1}{4} (J_4+l \cdot iD_4) e^{i l \alpha} {\bar s}^2
 \nonumber\\
 &&~~~~~~~~~~~~~~~~~~~
 \cdot
 t_{l,m+1}^{\dagger}({\bf k}) t_{l,m}({\bf k})
 \nonumber\\
 &+& h.c.
 \nonumber\\
 &-&
 \sum_{l=+,0,-} \sum_{\bf k} \sum_{m=1}^{6}
 \frac{1}{4} (J_4+l \cdot iD_4) e^{i l \alpha} {\bar s}^2
 \nonumber\\
 &&~~~~~~~~~~~~~~~~~~~
 \cdot
 t_{l,m+1}^{\dagger}({\bf k}) t_{-l,m}^{\dagger}(-{\bf k})
 \nonumber\\
 &+& h.c.
\end{eqnarray}
\end{subequations}
Every term in Eq. (\ref{eq:H_quad_n}) conserves the z-component of the total spin, $\sum_{i} {S}_{i,z}$, as the original Hamiltonian (\ref{eq:H_n}) does.

The quartic part of the Hamiltonian is decoupled in such a way that the resulting mean-field Hamiltonian conserves $\sum_{i} {S}_{i,z}$.
Readers interested in the details are referred to Appendix \ref{appendix:mean-field}.
The corresponding mean-field parameters are defined in (\ref{eq:mean_field_paramters}).
The resultant mean-field Hamiltonian can be arranged in the following form.
\begin{eqnarray}
 \mathcal{H}_{MF}
 &=&
 N_{uc} ( \epsilon_o + \epsilon_{PQ} )
 -
 \frac{1}{4}
 \sum_{l=+,0,-}
 \sum_{\bf k}
 \mathrm{Tr} {\bf M}_{l}({\bf k})
 \nonumber\\
 &+&
 \frac{1}{2}
 \sum_{l=+,0,-}
 \sum_{\bf k}
 \Lambda_{l}^{\dagger}({\bf k}) {\bf M}_{l}({\bf k}) \Lambda_{l}({\bf k}),
 \label{eq:H_MF}
\end{eqnarray}
where
\begin{eqnarray}
 \Lambda_{l}({\bf k})
 =
 \left[ t_{l,1}({\bf k}),\cdots, t_{l,6}({\bf k}), t_{-l,1}^{\dagger}(-{\bf k}),\cdots, t_{-l,6}^{\dagger}(-{\bf k}) \right]^{T},
 \nonumber\\
\end{eqnarray}
${\bf M}_l({\bf k})$ is a 12$\times$12 Hermitian matrix, 
and $\epsilon_{PQ}$ is the mean-field contribution to the energy and defined in (\ref{eq:epsilon_PQ}).
At this point, it is important to note that the mean-field Hamiltonian describes the triplon excitations and it is block diagonalized into three triplon modes of $l=+,0,-$.
As mentioned earlier, $l$ is the magnetic quantum number of the triplon, which reflects the SO(2) global spin rotation symmetry of the original and mean-field Hamiltonian.
This fact can also be seen from the following expression of the magnetization;
\begin{eqnarray}
  M_z
  &=&
  \frac{1}{6N_{uc}}\sum_{i} S_{i,z}
  \nonumber\\
  &=&
  \frac{1}{6N_{uc}}
  \sum_{l=+,0,-}
  \sum_{\bf k} \sum_{m=1}^{6} 
  l
  \cdot
  t_{l,m}^{\dagger}({\bf k}) t_{l,m}({\bf k}).
\end{eqnarray}

The mean-field Hamiltonian is diagonalized via the Bogoliubov transformation\cite{BlaizotRipka} which transforms $t$-bosons to bosonic quasiparticles ($\gamma$);
\begin{eqnarray}
 \Gamma_{l}({\bf k})
 = {\bf T}_{l}({\bf k}) \Lambda_{l}({\bf k}),
\end{eqnarray}
where
\begin{eqnarray}
 \Gamma_{l}({\bf k})
 =
 \left[ \gamma_{l,1}({\bf k}),\cdots, \gamma_{l,6}({\bf k}), \gamma_{-l,1}^{\dagger}(-{\bf k}),\cdots, \gamma_{-l,6}^{\dagger}(-{\bf k}) \right]^{T},
 \nonumber\\
\end{eqnarray}
and ${\bf T}_{l}({\bf k})$ is the transformation matrix.
The transformation matrix satisfies the following relation due to the bosonic statistics of the quasiparticles.
\begin{eqnarray}
  {\bf T}_{l}({\bf k}) {\bf I}_{B} {\bf T}_{l}^{\dagger}({\bf k}) = {\bf I}_{B},
  \label{eq:bosonic_constraint}
\end{eqnarray}
where ${\bf I}_{B}$ is the diagonal matrix with the first (second) six diagonal elements being $+ 1$ ($- 1$).
Because of this bosonic constraint, ${\bf I}_B {\bf M}_{l}({\bf k})$ is diagonalized instead of ${\bf M}_{l}({\bf k})$, so the eigenvalue problem has the following form;
\begin{eqnarray}
  {\bf T}_{l}({\bf k}) {\bf I}_B {\bf M}_l ({\bf k}) {\bf T}_{l}^{-1}({\bf k}) = {\bf I}_B {\boldsymbol \Omega}_{l}({\bf k}),
\end{eqnarray}
where ${\boldsymbol \Omega}_{l}({\bf k})$ is the diagonal matrix containing eigenvalues $\omega_{l,m}({\bf k}) ~ (m=1,\cdots,6)$ in the first and second half of its diagonal.
It must be noticed that ${\bf T}_{l}^{-1}({\bf k})$ contains eigenvectors of the eigenvalue problem in its columns and the eigenvectors are normalized according to Eq. (\ref{eq:bosonic_constraint}).
Diagonalizing the mean-field Hamiltonian with the above transformation leads to the following ground state energy per site.
\begin{eqnarray}
\epsilon_{gr} 
= 
\frac{1}{12} 
\left[ 
\epsilon_o 
+
\epsilon_{PQ}
-
\frac{1}{4N_{uc}}
\sum_{l=+,0,-}
\sum_{\bf k}
\mathrm{Tr} {\bf M}_{l} ({\bf k})
\right.
\nonumber\\
+
\left.
\frac{1}{2N_{uc}}
\sum_{l=+,0,-}
\sum_{\bf k}
\sum_{m=1}^{6}
\omega_{l,m} ({\bf k})
\right]
\end{eqnarray}
The energy is a function of the condensate density $\bar{s}$, Lagrange multiplier $\mu$, and mean-field parameters in (\ref{eq:mean_field_paramters}).
The ground state is determined by solving Eq. (\ref{eq:SC_eq_s_mu}) and (\ref{eq:mean_field_paramters}) self-consistently.
\begin{eqnarray}
\frac{\partial \epsilon_{gr}}{\partial \bar{s}} = 0,
~~
\frac{\partial \epsilon_{gr}}{\partial \mu} = 0.
\label{eq:SC_eq_s_mu}
\end{eqnarray}

\subsection{Ground state energy and triplon dispersion}

In this subsection, the results of the bond operator mean-field theory are presented.
We use the coupling constants estimated by the series expansion study\cite{neutron_series} for Rb$_2$Cu$_3$SnF$_{12}$;
\begin{eqnarray}
 &&J_2 = 0.95 ~ J_1,
 \nonumber\\
 &&J_3 = 0.85 ~ J_1,
 \nonumber\\
 &&J_4 = 0.55 ~ J_1,
 \nonumber\\
 &&d_z=D_n/J_n=0.18,
\end{eqnarray}
where $n=1,\cdots,4$.
For the DM interactions, increasing $d_z$ from 0 to 0.18, we investigate the effect of the DM interactions on the ground state energy and the triplon dispersions.

\begin{figure}[t]
 \centering
 \includegraphics[width=0.8\linewidth,angle=270]{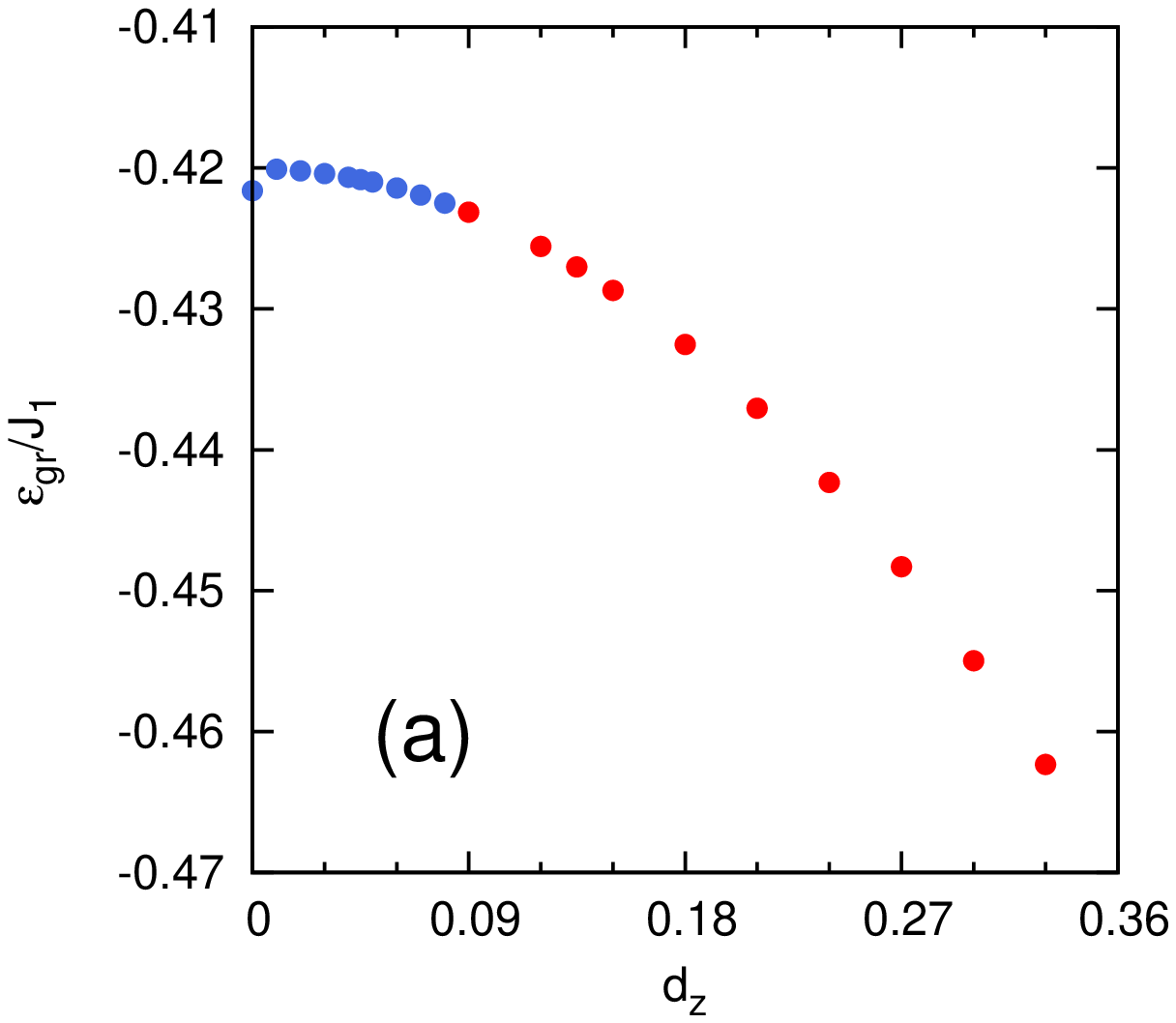}
 \includegraphics[width=0.8\linewidth,angle=270]{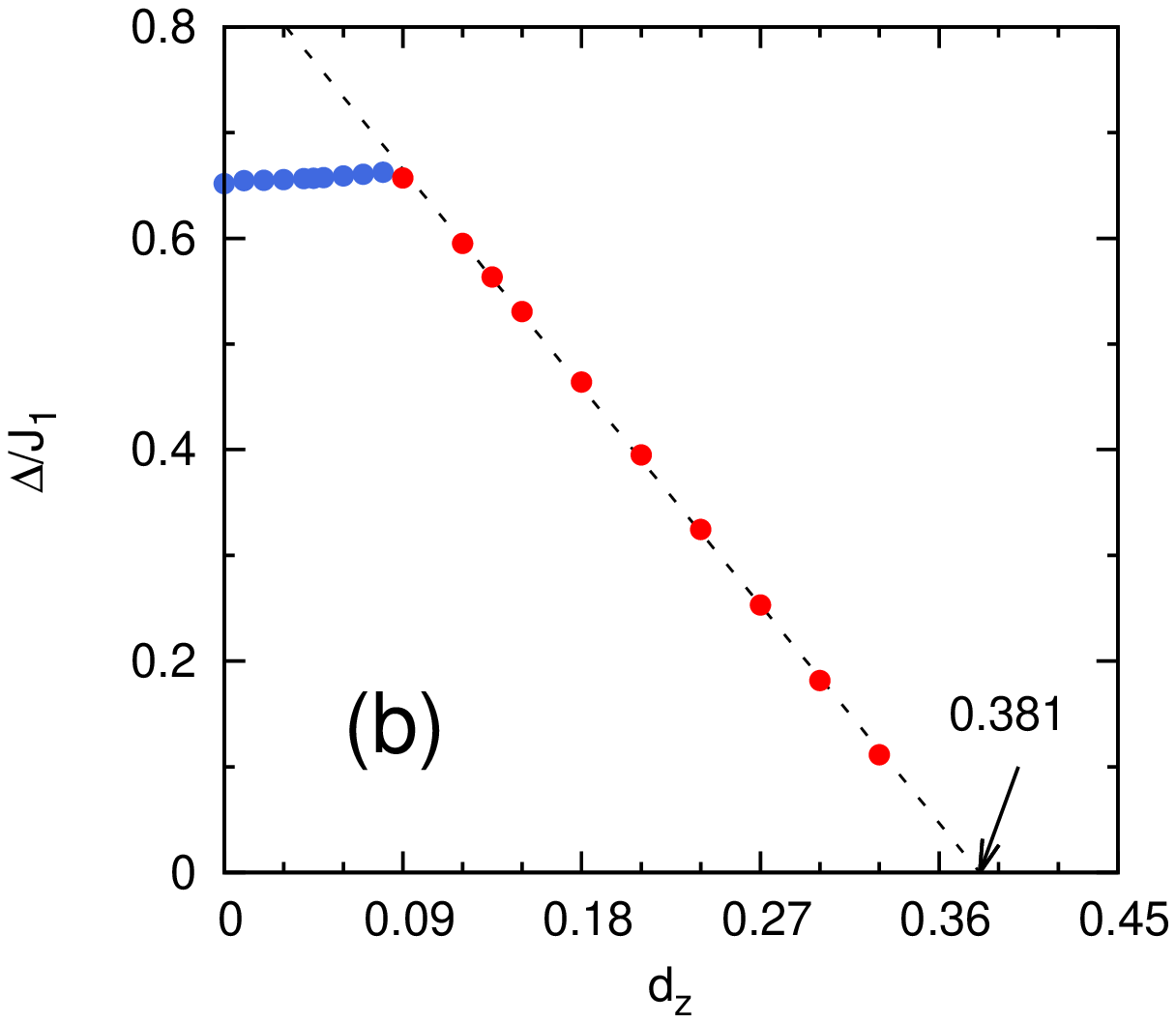}
 \caption{
 (Color online) Plots of (a) the ground state energy per site ($\epsilon_{gr}$) and (b) the spin gap ($\Delta$) versus the strength of the DM interaction ($d_z$).
 In the plots, the color represents the position of the spin gap;
 light blue for the spin gap at the K point and red for the spin gap at the $\Gamma$ point.
 The shift of the spin gap position from the K to $\Gamma$ point happens at $d_z=0.09$.
 In figure (b), the dashed line indicates the linear fitting for the spin gaps occurring at the $\Gamma$ point.
 According to the linear fitting, the spin gap is closed and a magnetic transition occurs at $d_{z,c}=0.381$.
 \label{fig:SC_sol}}
\end{figure}

The ground state energy is plotted for various values of $d_z$ in Fig. \ref{fig:SC_sol} (a).
As shown in the figure, the ground state energy decreases monotonically once we include the DM interactions.
The energy at $d_z=0$ is slightly off from the general tendency.
This is due to the fact that we employed different mean-field decouplings for the cases of $d_z=0$ and $d_z \ne 0$ 
because each case has different symmetry; SO(3) spin rotation symmetry when $d_z=0$ and SO(2) symmetry when $d_z \ne 0$.
How the mean-field decoupling is done in each case is explained in the Appendix \ref{appendix:mean-field}.
From the plot of the ground state energy, it turns out that the DM interaction tends to stabilize the valence bond state with the 12-site unit cell in the deformed kagome lattice antiferromagnet.
The ground state energy per site at $d_z=0.18$ is $-0.433 ~ J_1$.

\begin{table}
\begin{ruledtabular}
\begin{tabular}{lccc}
 & $\epsilon_{gr} / J_1$ & $\Delta / J_1 (= h_c/J_1)$ & $h_s/J_1$
 \\
 \hline
 BOT$_{quad}$ & -0.422 & 0.476 & $-$
 \\
 BOT$_{quartic}$ & -0.433 & 0.464 & $-$
 \\
 ED$_{12}$ & -0.453 & 0.371 & 2.809
 \\
 ED$_{24}$ & -0.448 & 0.229 & 2.809
 \\
 SC$_{24}$ & $-$ & 0.47 & 2.73
 \\
 SC$_{48}$ & $-$ & 0.47 & 2.73
 \\
 \hline
 Neutron-series\cite{neutron_series} & $-$ & 0.126 & $-$
\end{tabular}
\end{ruledtabular}
\caption{
The ground state energy per site ($\epsilon_{gr}$), the spin gap ($\Delta$), the critical magnetic field ($h_c$), and the saturation magnetic field ($h_s$) when $d_z=0.18$.
\label{tab:energy_spingap}}
\end{table}

\begin{figure*}
 \centering
 \includegraphics[width=\linewidth]{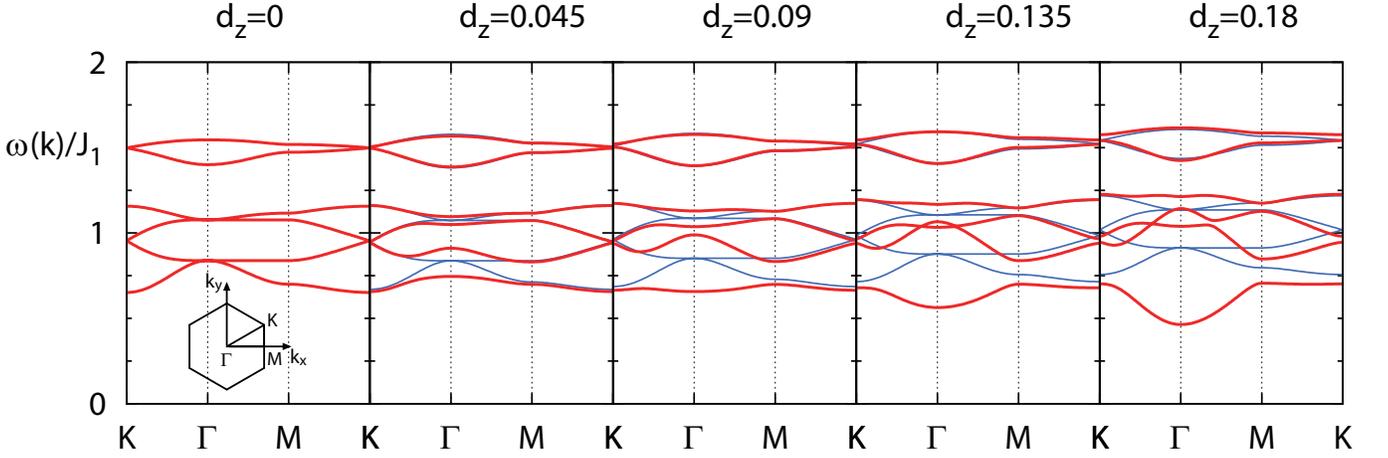}
 \caption{(Color online) Variation in the triplon dispersions caused by the DM interaction ($d_z$).
 Color indicates the magnetic quantum number of the triplon dispersion; thick red for $l=+,-$ and thin blue for $l=0$.
 When $d_z=0$, three modes with $l=+,0,-$ are all degenerate, so they are denoted with one color, red.
 The above plot shows that as the strength of the DM interaction increases, significant change occurs around the $\Gamma$ point so that the energy minimum (or the spin gap) shifts from the K to $\Gamma$ point at $d_z=0.09$.
 \label{fig:triplon_dispersion}}
\end{figure*}

Figure \ref{fig:triplon_dispersion} shows the variation in the triplon dispersions due to the change in the strength of the DM interaction.
When there is no DM interaction ($d_z=0$), the triplon dispersions with $l=+,0,-$ are all degenerate due to SO(3) spin rotation symmetry.
The spin gap happens at the K point in the first Brillouin zone.
These behaviors are also consistent with the triplon dispersions with $J_1=J_2=J_3=J$ and $J_4=0.5J$ of Ref. 32.
As we increase $d_z$ from zero, the spin rotation symmetry is reduced to SO(2) symmetry by the DM interaction, 
so the threefold degenerate dispersions are split into two parts (thick red and thin blue) as shown in Fig. \ref{fig:triplon_dispersion}.
In fact, the triplon dispersion with $l=0$ (thin blue) is not affected by the existence of the DM interaction except for a small overall shift.
For the others of $l=+,-$ (thick red), which are still degenerate, the degeneracies at the $\Gamma$ and K points are lifted by the DM interactions.
In particular, there is significant change around the $\Gamma$ point so that the energy minimum (or the spin gap) shifts from the K to $\Gamma$ point at $d_z=0.09$.
The spin gap is plotted as a function of $d_z$ in Fig. \ref{fig:SC_sol} (b), 
where color denotes the position of the spin gap; 
light blue for the spin gap at the K point and red for the spin gap at the $\Gamma$ point.
As shown in the figure, the spin gap does not change much when it is located at the K point.
However, after the spin gap shifts from the K to $\Gamma$ point at $d_z=0.09$, it decreases linearly as a function of $d_z$.
The spin gap ($\Delta$) at $d_z=0.18$ is $0.464 J_1$.
The triplon dispersions at $d_z=0.18$ in Fig. \ref{fig:triplon_dispersion} show qualitatively the same behavior as the excitation spectra obtained in the neutron scattering experiment;\cite{neutron_series} the spin gap at the $\Gamma$ point and the lowest energy band being doubly degenerate.
Thus, the triplon excitation spectrum is captured well in the bond operator mean-field theory.
The above features can be obtained just from the quadratic Hamiltonian (\ref{eq:H_quad_n}) without the quartic contributions.
The quartic contributions generate only numerical corrections without changing the qualitative behaviors.

\begin{figure}[b!]
 \centering
 \includegraphics[width=\linewidth]{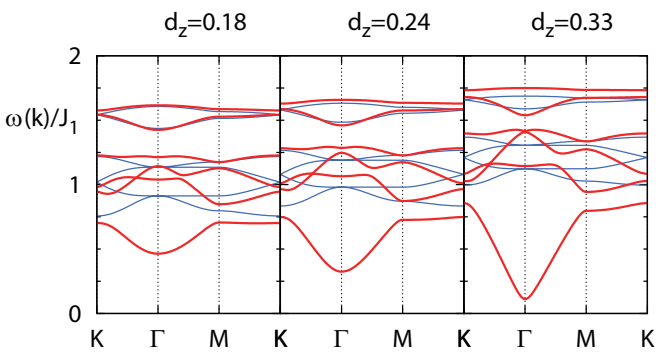}
 \caption{(Color online) Triplon dispersions beyond $d_z=0.18$.
 The lowest energy bands become linear around the $\Gamma$ point as we increase $d_z$.
 \label{fig:triplon_dispersion_2}}
\end{figure}

Now we consider the cases with $d_z > 0.18$.
If we increase $d_z$ beyond 0.18, the energy gap becomes smaller and the lowest energy bands become linear around the $\Gamma$ point as shown in Fig. \ref{fig:triplon_dispersion_2}.
Based on this fact, it is expected that the triplon dispersion becomes gapless at certain point of $d_{z,c}$.
The critical value $d_{z,c}=0.381$ is estimated by fitting linearly the spin gap values beyond $d_z=0.09$ as shown in Fig. \ref{fig:SC_sol} (b).
When the spin gap is closed, it leads to the condensation of the triplons at the K point and then causes a magnetic order in the system.
It is the transition to a magnetically ordered state induced by the Dzyaloshinskii-Moriya interaction.

Lastly, we comment on the behaviors of the VBS state and the triplon modes under the external magnetic field, ${\bf h}=h\hat{z}$.
The VBS state has no magnetization, thus the change due to the magnetic field occurs only to the triplon modes.
Under the magnetic field, the triplon modes with $l=+,-$ are coupled to the magnetic field while the other one with $l=0$ is not affected as shown in the following expression of the Zeeman interaction;
\begin{eqnarray}
 \mathcal{H}_h 
 &=&
 -h \hat{z} \cdot \sum_{i} {\bf S}_i
 \label{eq:H_h}
 \\
 &=&
 \sum_{l=+,0,-}
  \sum_{\bf k} \sum_{m=1}^{6} 
  (- l \cdot h)
  t_{l,m}^{\dagger}({\bf k}) t_{l,m}({\bf k}).
\end{eqnarray}
Now the triplon dispersions are modified as follows;
\begin{eqnarray}
 \omega_{l,m}({\bf k}) \rightarrow \omega_{l,m}({\bf k}) - l \cdot h,
\end{eqnarray}
where $\omega_{l,m}({\bf k}) ~ (l=+,0,-;m=1,\cdots,6)$ is the triplon dispersion when $h=0$.
As we increase $h$, the lowest band with $l=+$ mode hits the zero energy at the critical magnetic field $h_c = \Delta$,
where $\Delta$ is the spin gap when $h=0$.
Beyond $h_c$, the VBS state becomes unstable due to the gapless triplon mode with $l=+$, the condensation of which causes a magnetic ordering in the system.

The critical magnetic field, or the spin gap, is 0.464$J_1$ when $d_z=0.18$.
Comparing with the experimental result $\Delta_{exp}=0.126 J_1$,\cite{neutron_series} the critical field is overestimated in the bond operator mean-field theory.
For more accurate estimation of the spin gap, we conduct the exact diagonalization and strong coupling expansion studies in the next sections.
In those studies, we also investigate the magnetization process.

\section{EXACT DIAGONALIZATION\label{sec:ED}}

\begin{figure}
 \centering
 \includegraphics[width=0.7\linewidth]{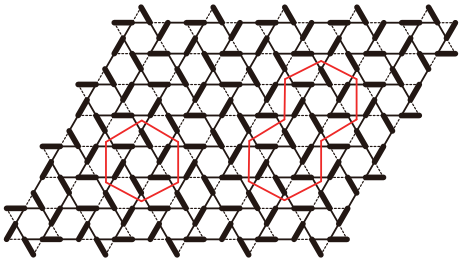}
 \caption{(Color online) Clusters with 12 and 24 sites used for the exact diagonalization study.\label{fig:ED_clusters}}
\end{figure}

\begin{figure}[b!]
 \centering
 \includegraphics[width=0.75\linewidth]{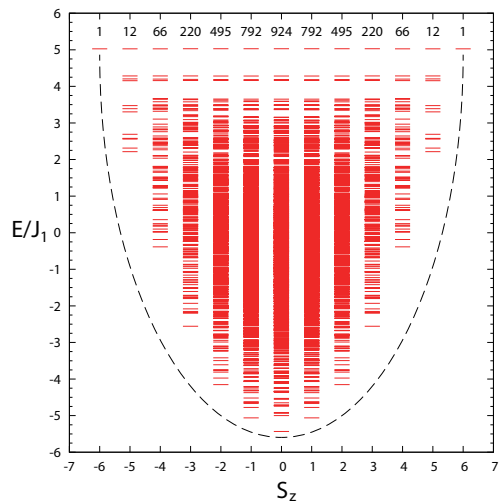}
 \caption{(Color online) Full energy spectrum of the 12-site cluster when there is no magnetic field.
 Dashed line provides a guide for the shape of the the function $E_o(S_z)$.
 \label{fig:ED_spectrum_12}}
\end{figure}

The deformed kagome lattice has the large unit cell of 12 sites.
Because of the large-size unit cell, it is difficult to study large cluster sizes for the exact diagonalization (ED) study.
Thus, we are limited to small size clusters shown in Fig. \ref{fig:ED_clusters} with periodic boundary condition.
In each cluster, we diagonalize the Hamiltonian
\begin{eqnarray}
 \mathcal{H}_J+\mathcal{H}_{D}+\mathcal{H}_h,
 \label{eq:H_for_ED_SC}
\end{eqnarray}
each term of which was already introduced in Eq. (\ref{eq:H_J_H_DM}), (\ref{eq:H_n}), and (\ref{eq:H_h}).
Fig. \ref{fig:ED_spectrum_12} shows the full spectrum of the 12-site cluster when there is no magnetic field.
Its ground state occurs in $S_z=0$ sector with the energy per site of -0.453$J_1$, where $S_z=\sum_{i=1}^{N_s} S_{i,z}$ and $N_s$ is the number of sites.
The spin gap, which is the energy difference between the ground state and the lowest energy state in $S_z=1$, is 0.371$J_1$.
These values are modified in 24-site cluster; the ground state energy per is -0.448$J_1$ and the spin gap is 0.229$J_1$.
It is interesting to note that the value of spin gap is not far from that estimated in the experiment, 0.126 $J_1$, even though it is obtained in the small cluster of 24 sites.

\begin{figure}
 \centering
 \includegraphics[width=0.75\linewidth,angle=270]{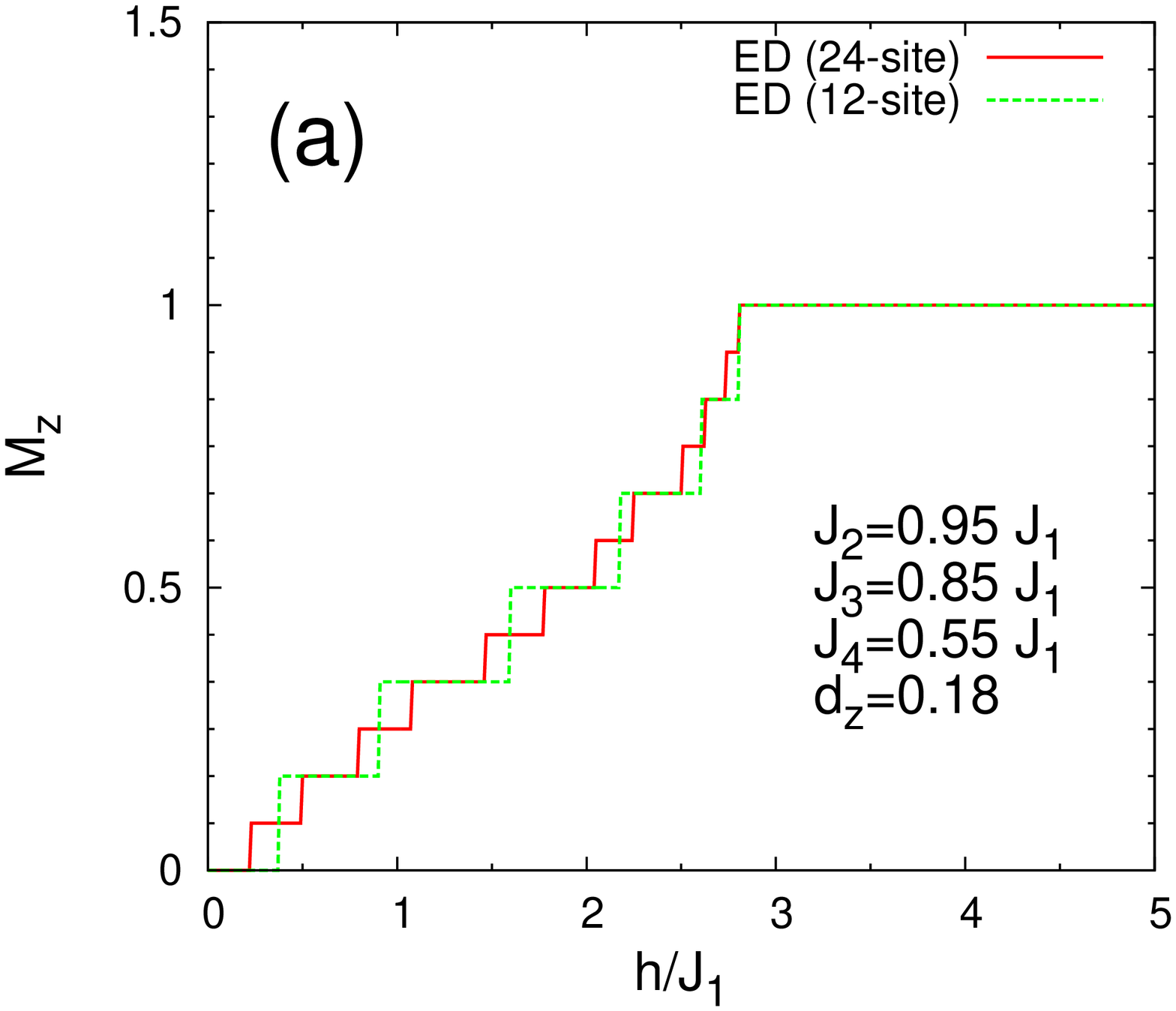}
 \includegraphics[width=0.75\linewidth,angle=270]{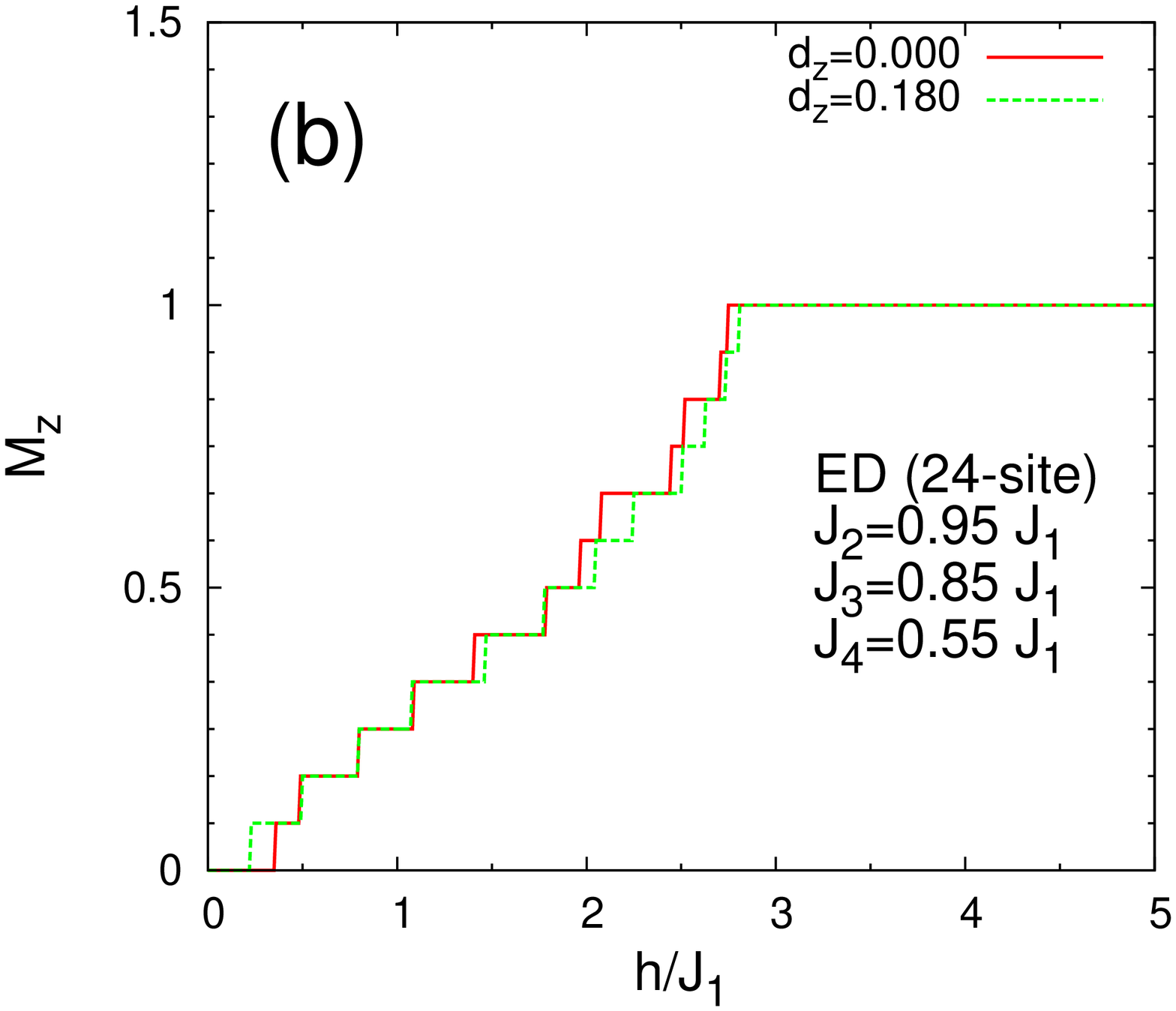}
 \caption{(Color online) Magnetization curves for the clusters with 12 and 24 sites.\label{fig:ED_magnetization}}
\end{figure}

If we include the Zeeman interactions and change the magnetic field ($h$), the ground state selected by the system changes.
The ground state is determined by the competition among the lowest energy states in the sectors of $S_z=0,\cdots,N_s/2$; we call these states $\left| E_o(S_z) \right>$.
When $h$ is small, the lowest energy state of $S_z=0$, $\left| E_o(S_z=0) \right>$, remains as the ground state.
The lowest one of $S_z=N_s/2$, $\left| E_o(S_z=N_s/2) \right>$, becomes the ground state when $h$ becomes large enough.
How the system moves through the lowest energy manifold $\left\{ \left| E_o(S_z) \right> | S_z=0,\cdots,N_s/2 \right\}$ in intermediate region of $h$ determines the profile of magnetization curve.
The energies of those states, $E_o(S_z)$, consist of concave curve in $E_o$ vs. $S_z$ plot as guided by the dashed line in Fig. \ref{fig:ED_spectrum_12}.
In that case, the system goes through all of the lowest energy states.\cite{Shunsuke}
As a result, the magnetization ($M_z=2S_z/N_s$) increases gradually from zero to one (see Fig. \ref{fig:ED_magnetization} (a)).
This can be understood by noting that the ground state is determined by the following equation;
\begin{eqnarray}
 \frac{\partial}{\partial S_z} \left[ E_o(S_z) - h \cdot S_z \right] = 0
\end{eqnarray}
If $E_o(S_z)$ is concave like that of 12-site cluster, then the minimum of $E_o(S_z) - h \cdot S_z$ moves gradually from $S_z=0$ to $S_z=N_s/2$.
If the energy curve of $\left\{ \left| E_o(S_z) \right> | S_z=0,\cdots,N_s/2 \right\}$ were convex, the ground state would jump abruptly from $\left| E_o(S_z=0) \right>$ to $\left| E_o(S_z=N_s/2) \right>$.
The gradual increase of magnetization from zero is consistent with the experimental result.\cite{suscep_magnetization_gap}

Now, we discuss about the effect of the DM interaction on the magnetization profile.
Fig. \ref{fig:ED_magnetization} (b) shows magnetizations for two cases of $d_z=0$ and 0.18 in 24-site cluster.
A few changes are observed.
First, the spin gap or the critical magnetic field ($\Delta=h_c$) is significantly decreased by the DM interaction; from 0.350 $J_1$ to 0.229 $J_1$.
This fact is consistent with the result of the bond operator mean-field theory.
DM interaction increases the saturation magnetic field $h_s$, at which the system becomes completely magnetized, as shown in Fig. \ref{fig:ED_magnetization} (b).
As a result, it extends the range, $h_s-h_c$, where the magnetization process goes on.

\section{STRONG COUPLING EXPANSION\label{sec:SCE}}

In this section, we derive an effective Hamiltonian when the magnetic field is strong enough to cause magnetization in the system.
It is used to investigate the magnetization process of the deformed kagome lattice antiferromagnet.
The effective Hamiltonian is obtained by applying the degenerate perturbation theory to the original Hamiltonian.
This approach is called the strong coupling expansion.\cite{Momoi_Totsuka}

We first arrange the Hamiltonian (\ref{eq:H_for_ED_SC}) as follows;
\begin{eqnarray}
 \mathcal{H}_J+\mathcal{H}_{D}+\mathcal{H}_h
 =
 \mathcal{H}_o + \mathcal{H}',
\end{eqnarray}
where
\begin{eqnarray}
 \mathcal{H}_o&=&\mathcal{H}_{J_1}+\mathcal{H}_{D_{1}}-\Delta_{dimer} \sum_{i} S_{i,z},
 \\
 \mathcal{H}'&=&\sum_{n=2}^{4} \left( \mathcal{H}_{J_n}+\mathcal{H}_{D_{n}} \right) +(\Delta_{dimer}-h) \sum_{i} S_{i,z}.
\end{eqnarray}
In the above equations, $\mathcal{H}_{J_n}$ is the Heisenberg interaction part with the coupling constant $J_n$ and $\mathcal{H}_{D_n}$ is the DM interaction part with $D_n$.
$\Delta_{dimer}=\epsilon_{t_{+}}-\epsilon_s$, where $\epsilon_{t_{+}}$ and $\epsilon_s$ are from Eq. (\ref{eq:H_dimer_energies}).
$\mathcal{H}_o$ is regarded as the unperturbed Hamiltonian, which provides proper basis to describe the magnetization process, and $\mathcal{H}'$ is the perturbation for $\mathcal{H}_o$.
Notice that $\mathcal{H}_o$ corresponds to the Hamiltonian of independent dimers under the magnetic field of $h=\Delta_{dimer}$, the eigenstates of which are given by direct product of dimer states in Eq. (\ref{eq:BOT_basis}).
The magnetic field of $h=\Delta_{dimer}$ in $\mathcal{H}_o$ is intended to make the states with $\epsilon_{t_{+}}$ and $\epsilon_s$ to have the degenerate ground state energy in each dimer.
Then, $\mathcal{H}_o$ has the degenerate ground subspace;
$
 V_o
 =
 \left\{
 \otimes_{i} | \psi \rangle_i ~ | ~ | \psi \rangle = | s \rangle, | t_{+} \rangle
 \right\}.
$
The other states are gapped out from the ground state subspace by $\Delta_{dimer}$.

\begin{figure}
 \centering
 \includegraphics[width=0.7\linewidth]{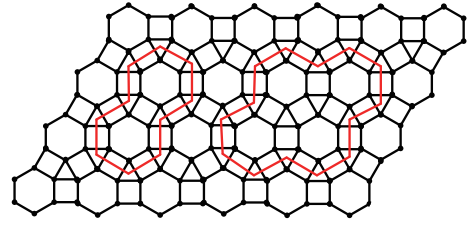}
 \caption{(Color online) Lattice structure for the effective hardcore boson model in the strong coupling expansion study.
 The finite clusters enclosed by the red line are used to solve the hardcore boson model exactly.
 In the lattice, each point denotes a dimer in the original lattice of spins in Fig. \ref{fig:dimer_configuration}.
 The left cluster in the figure is called 24-site cluster and the right 48-site cluster, referring to the number of sites in the original lattice.
 \label{fig:lattice_HCB}}
\end{figure}

The effective Hamiltonian is generated by applying the first-order perturbation theory on the degenerate subspace $V_o$.
The Hamiltonian can be mapped into the hardcore boson model.
To this end, we employ $b$-boson at each dimer and map the states $\left| s \right>$ and $\left| t_{+} \right>$ into zero-boson and one-boson states as follows;
\begin{subequations}
\begin{eqnarray}
 b^{\dagger} &=& \left| t_{+} \right> \left< s \right|,
 \\
 b &=& \left| s \right> \left< t_{+} \right|, 
 \\
 n &=& \left| t_{+} \right> \left< t_{+} \right|,
\end{eqnarray}
\end{subequations}
where $b^{\dagger}$ and $b$ are operators to create and annihilate a boson respectively and $n$ is the number operator for the bosons.
For the hardcore bosons, the original lattice in Fig. \ref{fig:dimer_configuration} is transformed to the lattice shown in Fig. \ref{fig:lattice_HCB}; 
the dimers in the former are mapped into the points in the latter, so that the 12-site unit cell in the former becomes the 6-point hexagon in the latter.
The first order effective Hamiltonian is written as follows in terms of the hardcore bosons.
\begin{eqnarray}
 && \mathcal{H}_{eff}
 \nonumber\\
 &=&
 N_{uc} \cdot 6 \epsilon_s
 \nonumber\\
 &+& 
 \frac{1}{4} (J_2+iD_2)
 \sum_{\bf r} \sum_{m=1}^{6}
 b_m^{\dagger}({\bf r}) b_{m+1}({\bf r})
 \nonumber\\
 &+&h.c.
 \nonumber\\
 &+& 
 \frac{1}{4} (J_3+iD_3)
 \sum_{\bf r} \sum_{(m,n;{\bf R}) \in I_3}
 b_m^{\dagger}({\bf r}) b_n({\bf r}+{\bf R})
 \nonumber\\
 &+&h.c.
 \nonumber\\
 &-& 
 \frac{1}{4} (J_4+iD_4) e^{i \alpha}
 \sum_{\bf r} \sum_{m=1}^{6}
 b_{m+1}^{\dagger}({\bf r}) b_m({\bf r})
 \nonumber\\
 &+&h.c.
 \nonumber\\
 &+&
 \frac{1}{4} J_2
 \sum_{\bf r} \sum_{m=1}^{6}
 n_m({\bf r}) n_{m+1}({\bf r})
 \nonumber\\
 &+&
 \frac{1}{4} J_3
 \sum_{\bf r} \sum_{(m,n;{\bf R}) \in I_3}
 n_m({\bf r}) n_n({\bf r}+{\bf R})
 \nonumber\\
 &+&
 \frac{1}{4}  J_4
 \sum_{\bf r} \sum_{m=1}^{6}
 n_{m+1}({\bf r}) n_m({\bf r})
 \nonumber\\
 &+&
 (\Delta_{dimer}-h)
 \sum_{\bf r} \sum_{m=1}^{6}
 n_m({\bf r}),
 \label{eq:H_eff}
\end{eqnarray}
where $b_7=b_1$ and $n_7=n_1$.
The  hardcore boson model Hamiltonian consists of the hopping and nearest neighbor repulsion.
In this effective Hamiltonian, the hopping amplitude is comparable to the repulsion energy as shown in Eq. (\ref{eq:H_eff}).
This fact makes it difficult to predict the ground state of the Hamiltonian.
For this reason, the Hamiltonian is diagonalized exactly for two clusters as shown in Fig. \ref{fig:lattice_HCB}.
The cluster shown in the left of the figure is the 24-site cluster and the 48-site cluster is shown in the right.
These clusters are used for the exact diagonalization studies of the effective Hamiltonian.
Since we are studying the hardcore boson model, the total number of bosons is allowed to have some number between zero and the number of sites in the underlying cluster.
To obtain the magnetization profile, we find the ground state at each magnetic field.
In the hardcore boson model, the magnetic field controls the number of bosons like the chemical potential.
If we increase the magnetic field, then the number of bosons in the ground state increases.
In fact, the number of boson is nothing but the magnetization of the ground state as the following relation indicates;
\begin{eqnarray}
 M_z = \frac{1}{6N_{uc}} \sum_{\bf r} \sum_{m=1}^{6} n_m({\bf r}).
\end{eqnarray}

\begin{figure}
 \centering
 \includegraphics[width=0.75\linewidth,angle=270]{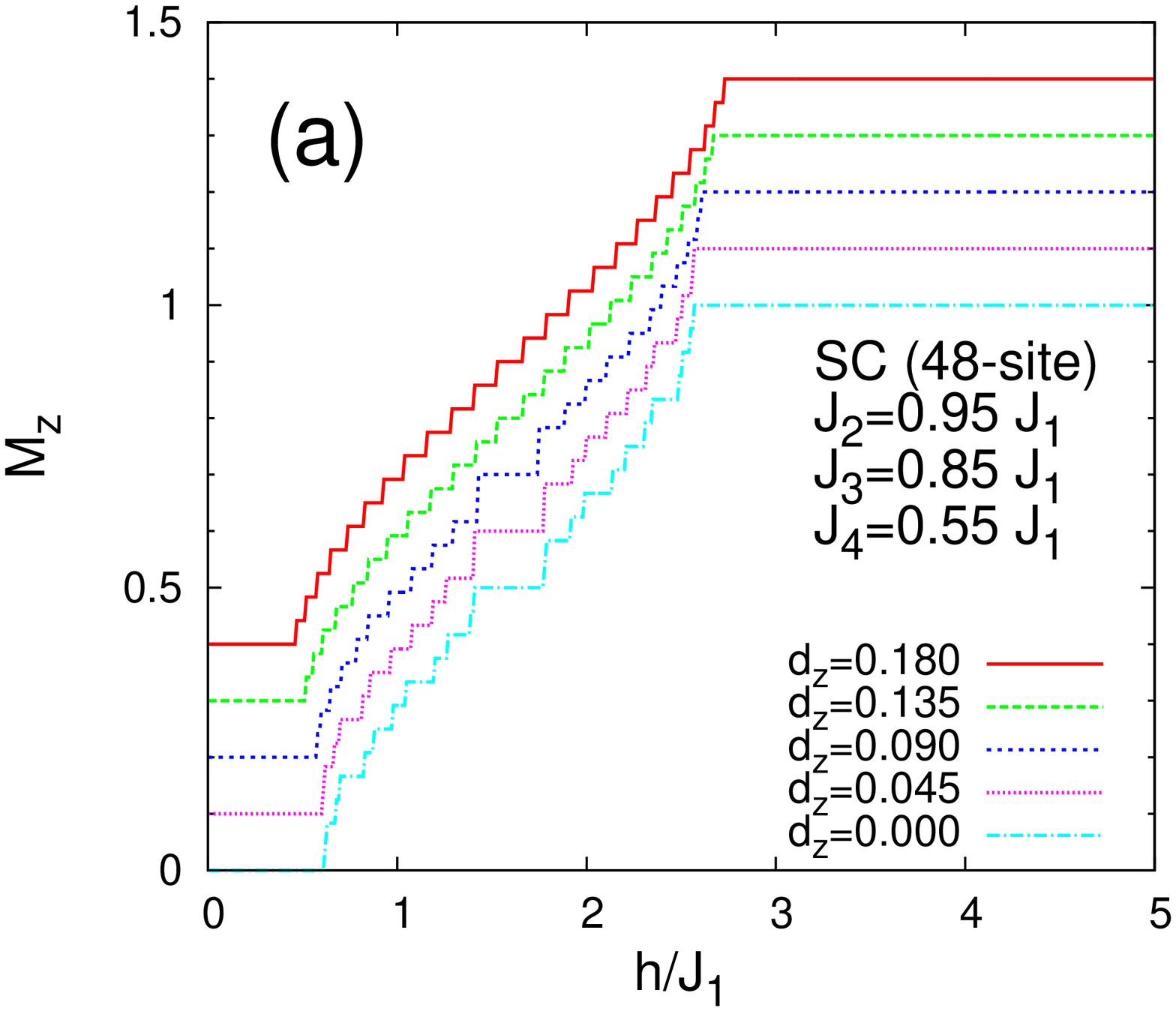}
 \includegraphics[width=0.75\linewidth,angle=270]{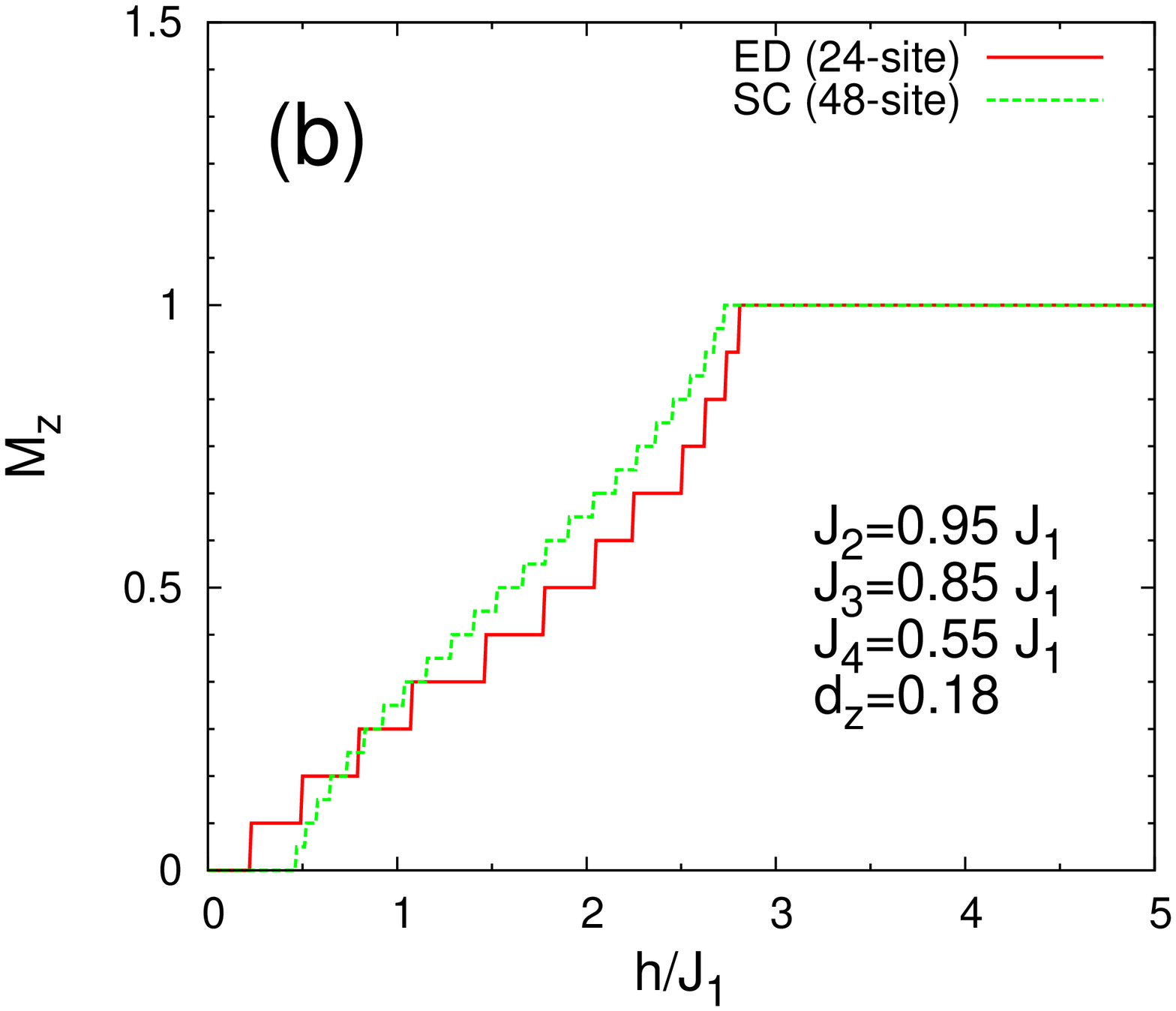}
 \caption{
 (Color online) 
 Magnetization curves obtained in the strong coupling expansion study.
 In figure (a), the magnetization curve for each value of $d_z$ is shifted upward to avoid overlapping of the curves.
 Each curve starts from $M_z=0$ at $h=0$.
 \label{fig:SC_M_z}}
\end{figure}

The magnetization of the 48-site cluster is plotted in Fig. \ref{fig:SC_M_z} (a) for several values of $d_z$.
Interestingly, the magnetization plateau at $M_z=1/2$ appears when the DM interaction is weak ($d_z = 0 \sim 0.09$).
The magnetization plateau at $M_z=1/2$ becomes narrower and narrower and eventually vanishes as the strength of the DM interaction increases from $d_z=0$ to 0.18.
The same behavior in magnetization is observed in the 24-site cluster as well.

Although $M_z=1/2$ plateau is an interesting point of the strong coupling expansion study, we must be careful about the interpretation.
First of all, the exact diagonalization study does not show any magnetization step which can be considered as a plateau.
On the other hand, the effective Hamiltonian in the strong coupling expansion is obtained within the first order perturbation theory, so there are many higher order interactions which are ignored.
In this regard, higher order interactions are to be included in the effective Hamiltonian in order to describe the magnetization process in terms of the hardcore bosons more accurately.
Nonetheless, the first order effective Hamiltonian captures the overall increasing behavior of the magnetization curve, already found in the exact diagonalization study (see Fig. \ref{fig:SC_M_z} (b)).
The effect of the DM interaction on the magnetization is also consistent with the exact diagonalization result;
the critical field value $h_c$ is decreased and the saturation field strength $h_s$ is increased when we increase the strength of DM interaction.
It turns out that $h_c = 0.47 J_1 $ and $h_s = 2.73 J_1$ for both of the 24- and 48-site clusters at $d_z=0.18$.
The critical and saturation field strengths from various calculations are listed in Table \ref{tab:energy_spingap}.

\section{DISCUSSION\label{sec:DISCUSSION}}

We investigated the effect of the DM interaction in a model appropriate for Rb$_2$Cu$_3$SnF$_{12}$ via various theoretical approaches.
Assuming the DM vectors are perpendicular to the lattice plane, we consider the valence bond solid state with the 12-site unit cell as the ground state of the system, which was confirmed by the neutron scattering experiment.
First, using the bond operator mean-field theory, 
we study the effect of the DM interactions on the VBS state with the 12-site unit cell and the excitation spectrum.
It is shown that the ground state becomes more stabilized by the DM interactions.
The DM interactions have significant effect on the excitation spectrum or the triplon dispersions.
First of all, the DM interaction breaks the SO(3) spin symmetry down to the SO(2) spin rotation within the lattice plane.
As a result, the triplon modes with $l=+,-$, which are still degenerate, are separated from the $l=0$ mode,
with significant modification of the spectrum around the $\Gamma$ point by the DM interaction.
This change induces the shift of the spin gap (the minimum triplon gap) position from the K to $\Gamma$ point in the first Brillouin zone at $d_z=0.09$.
The spin gap at the $\Gamma$ point and shapes of the lowest three triplon dispersions at $d_z=0.18$ in Fig. \ref{fig:triplon_dispersion} are characteristic features observed in the neutron scattering experiment with Rb$_2$Cu$_3$SnF$_{12}$.
Therefore, the bond operator theory correctly captures the DM interaction effect in Rb$_2$Cu$_3$SnF$_{12}$.

In the deformed kagome lattice antiferromagnet, we can think about two kinds of transitions induced by the Dzyaloshinkii-Moriya interaction and magnetic field respectively.
For the former transition, the transition point is predicted to be $d_{z,c}=0.381$ by fitting linearly the spin gaps for $d_z$ bigger than 0.09 in the bond operator theory.
At the transition point, the gapless triplon mode with $l=+,-$ are condensed to cause the magnetic order.
At this transition, the SO(2) spin symmetry is spontaneously broken.
In fact, there were several studies on such kind of transition and possible magnetic ordering for spin-1/2 antiferromagnet on the uniform kagome lattice.\cite{DM_kagome,DM_ED,DM_SB_Messio,DM_SB_Hur}
Among those studies, the exact diagonalization study showed that the critical value for the magnitude of the DM vector is $D/J = 0.1$, where $J$ ($D$) is the coupling constant of the Heisenberg (DM) interactions in the uniform kagome lattice.\cite{DM_ED}
Beyond the critical value, the system enters the N\'eel ordered state.
Comparing our results with their study, the magnetic transition induced by the Dzyaloshinkii-Moriya interaction in the deformed kagome lattice antiferromagnet seems to have similar structure but with a larger value for the critical DM interaction.
However, it should be noted that the bond operator theory tends to overestimate the spin gap.
Comparing with the neutron scattering experiment result, the spin gap value from the bond operator theory is about four times larger than the experimental result (see Table \ref{tab:energy_spingap}).
Considering the linearly decreasing behavior of the spin gap with respect to the DM interaction in the deformed kagome lattice antiferromagnet,
the critical value $d_{z,c}$ is also expected to be overestimated.
The critical point can be estimated to be $d_{z,c}=0.23$ by using the following fact and assumption;
(i) the fact that the bond operator theory prediction for the spin gap is four times larger than the experimental value at $d_z=0.18$,
and
(ii) the assumption that the spin gap of Rb$_2$Cu$_3$SnF$_{12}$ is a linear function of $d_z$ and it has the same slope that is predicted in the bond operator theory.
According to this estimation, Rb$_2$Cu$_3$SnF$_{12}$ is located pretty close to the critical point.
This situation is very similar to one of the Herbertsmithite ZnCu$_3$(OH)$_6$Cl$_2$.
The strength of DM interaction in the Herbertsmithite was measured to be $D/J \sim 0.08$ via electron spin resonance experiment,\cite{DM_ESR_Zorko_2008} 
which is pretty close to its critical value $D/J = 0.1$ from exact diagonalization study.\cite{DM_ED}
In this regard, Rb$_2$Cu$_3$SnF$_{12}$ and ZnCu$_3$(OH)$_6$Cl$_2$ have common feature of being close to the quantum phase transition induced by the Dzyaloshinskii-Moriya interactions even though they have different lattice symmetries.

The other transition occurs via the magnetization process.
Under magnetic field perpendicular to the lattice plane, the lowest $l=+$ triplon mode becomes gapless at the critical magnetic field $h_c=\Delta$.
Then, it is condensed to induce the magnetization along the direction perpendicular to the lattice plane.
The entire magnetization process is studied by both exact diagonalization and strong coupling expansion.
In particular, the exact diagonalization is used to estimate more accurate magnitude of the spin gap.

From the exact diagonaliztion study, the spin gap was computed to be $0.229 J_1$ for the 24-site cluster with the periodic boundary condition.
This value is not far from the experimental result, $0.126 J_1$.
The system exhibits continuously increasing magnetization curve without any magnetization plateau regardless of the DM interaction strength.
The DM interaction reduces the critical field value $h_c$ and increases the saturation field strength $h_s$.
As a result, it extends the range, $h_s-h_c$, where magnetization process goes on.

In the strong coupling expansion, we restrict the Hilbert space to the subspace of relavent states for the magnetization process and then derive an effective Hamiltonian by applying the first order perturbation theory to the subspace.
The effective Hamiltonian is mapped to the hardcore boson model, which consists of the hoppings and repulsions of the hadrcore bosons under  ``chemical potential'' given by the magnetic field. 
Exact calculations on several finite clusters at $d_z=0.18$ show qualitatively the same magnetization curve compared to the exact diagonalization result of the full Hamiltonian.
Interestingly, $M_z=1/2$ magnetization plateau was observed.
It is concluded, however, that the effective Hamiltonian has to include higher order interactions to describe the magnetization process accurately.

\acknowledgements
This work was supported by the NSERC of Canada and Canadian Institute for Advanced Research (KH, YBK) and the National Research Foundation of Korea (NRF) funded by the Korea government (MEST) through the Quantum Metamaterials Research Center, No. 2008-0062238 (KP).

\appendix
\section{Mean-field decoupling\label{appendix:mean-field}}
In this appendix, we explain the mean-field decoupling of the quartic part of the Hamiltonian (\ref{eq:BOT_Hamiltonian}).
The quartic part can be arranged in following form.
\begin{widetext}
\begin{eqnarray}
 &&
 H_{quartic}
 \nonumber\\
 &=&
 \sum_{\bf r}
 \sum_{(m,n)}
 A_{mn}
 \left[ -t_{+,m}^{\dagger}({\bf r}) t_{0,m}({\bf r}) + t_{0,m}^{\dagger}({\bf r}) t_{-,m}({\bf r}) \right]
 \left[ -t_{0,n}^{\dagger}({\bf r}+{\bf R}_{mn}) t_{+,n}({\bf r}+{\bf R}_{mn}) + t_{-,n}^{\dagger}({\bf r}+{\bf R}_{mn}) t_{0,n}({\bf r}+{\bf R}_{mn}) \right]
 \nonumber\\
 &+&h.c.
 \nonumber\\
 &+&
 \sum_{\bf r}
 \sum_{(m,n)}
 B_{mn}
 \left[ t_{+,m}^{\dagger}({\bf r}) t_{+,m}({\bf r}) - t_{-,m}^{\dagger}({\bf r}) t_{-,m}({\bf r}) \right]
 \left[ t_{+,n}^{\dagger}({\bf r}+{\bf R}_{mn}) t_{+,n}({\bf r}+{\bf R}_{mn}) - t_{-,n}^{\dagger}({\bf r}+{\bf R}_{mn}) t_{-,n}({\bf r}+{\bf R}_{mn}) \right],
 \nonumber\\
\label{eq:H_quartic}
\end{eqnarray}
\end{widetext}
where $(m,n)$, ${\bf R}_{mn}$, $A_{mn}$, and $B_{mn}$ are listed in Table \ref{tab:R_A_B}.

\begin{table}
\begin{tabular}{cc|c|c|c}
\hline
$m$ & $n$ & ${\bf R}_{mn}$ & $A_{mn}$ & $B_{mn}$
\\
\hline
1 & 2 & ${\bf 0}$ & $\frac{1}{4}(J_2+iD_2)$ & $\frac{1}{4}J_2$
\\
2 & 3 & ${\bf 0}$ & &
\\
3 & 4 & ${\bf 0}$ & &
\\
4 & 5 & ${\bf 0}$ & &
\\
5 & 6 & ${\bf 0}$ & &
\\
6 & 1 & ${\bf 0}$ & &
\\
\hline
1 & 5 & ${\bf r}_C$ & $\frac{1}{4}(J_3+iD_3)$ & $\frac{1}{4}J_3$
\\
3 & 1 & $-{\bf r}_B$ & &
\\
5 & 3 & ${\bf r}_A$ & &
\\
2 & 6 & $-{\bf r}_A$ & &
\\
4 & 2 & $-{\bf r}_C$ & &
\\
6 & 4 & ${\bf r}_B$ & &
\\
\hline
1 & 6 & ${\bf 0}$ & $\frac{1}{4}(J_4+iD_4) e^{i \alpha}$ & $\frac{1}{4}J_4$
\\
2 & 1 & ${\bf 0}$ & &
\\
3 & 2 & ${\bf 0}$ & &
\\
4 & 3 & ${\bf 0}$ & &
\\
5 & 4 & ${\bf 0}$ & &
\\
6 & 5 & ${\bf 0}$ & &
\\
\hline
\end{tabular}
\caption{$(m,n)$, ${\bf R}_{mn}$, $A_{mn}$, and $B_{mn}$ in (\ref{eq:H_quartic}).
\label{tab:R_A_B}}
\end{table}

We decouple the quartic terms in such a way that the resultant mean-field Hamiltonian preserves the SO(2) spin rotation symmetry.
To this end, we define the following mean-field parameters;
\begin{subequations}
\begin{eqnarray}
 && P_{mn}^{++}=\langle t_{+,m}({\bf r}) t_{+,n}^{\dagger}({\bf r}+{\bf R}_{mn}) \rangle,
 \\
 && P_{mn}^{--}=\langle t_{-,m}({\bf r}) t_{-,n}^{\dagger}({\bf r}+{\bf R}_{mn}) \rangle,
 \\
 && P_{mn}^{00}=\langle t_{0,m}({\bf r}) t_{0,n}^{\dagger}({\bf r}+{\bf R}_{mn}) \rangle,
 \\
 && Q_{mn}^{+-}=\langle t_{+,m}({\bf r}) t_{-,n}({\bf r}+{\bf R}_{mn}) \rangle,
 \\
 && Q_{mn}^{-+}=\langle t_{-,m}({\bf r}) t_{+,n}({\bf r}+{\bf R}_{mn}) \rangle,
 \\
 && Q_{mn}^{00}=\langle t_{0,m}({\bf r}) t_{0,n}({\bf r}+{\bf R}_{mn}) \rangle.
\end{eqnarray}
 \label{eq:mean_field_paramters}
\end{subequations}
The above parameters are defined for $(m,n)$ listed in Table \ref{tab:R_A_B} and assumed to be independent of the lattice vector ${\bf r}$.
With these mean-field parameters, the quartic part of the Hamiltonian is decoupled into the following form.
\begin{eqnarray}
 && H_{quartic}
 \nonumber\\
 &=&
 N_{uc} \cdot \epsilon_{PQ}
 \nonumber\\
 &+&
 \sum_{\bf k} \sum_{(m,n)}
 e^{i {\bf k} \cdot {\bf R}_{mn}} a_{mn}^{++} \cdot t_{+,m}^{\dagger}({\bf k}) t_{+,n}({\bf k}) + h.c.
 \nonumber\\
 &+&
 \sum_{\bf k} \sum_{(m,n)}
 e^{i {\bf k} \cdot {\bf R}_{mn}} a_{mn}^{--} \cdot t_{-,m}^{\dagger}({\bf k}) t_{-,n}({\bf k}) + h.c.
 \nonumber\\
 &+&
 \sum_{\bf k} \sum_{(m,n)}
 e^{i {\bf k} \cdot {\bf R}_{mn}} a_{mn}^{00} \cdot t_{0,m}^{\dagger}({\bf k}) t_{0,n}({\bf k}) + h.c.
 \nonumber\\
 &+&
 \sum_{\bf k} \sum_{(m,n)}
 e^{i {\bf k} \cdot {\bf R}_{mn}} b_{mn}^{+-} \cdot t_{+,m}^{\dagger}({\bf k}) t_{-,n}^{\dagger}(-{\bf k}) + h.c.
 \nonumber\\
 &+&
 \sum_{\bf k} \sum_{(m,n)}
 e^{i {\bf k} \cdot {\bf R}_{mn}} b_{mn}^{-+} \cdot t_{-,m}^{\dagger}({\bf k}) t_{+,n}^{\dagger}(-{\bf k}) + h.c.
 \nonumber\\
 &+&
 \sum_{\bf k} \sum_{(m,n)}
 e^{i {\bf k} \cdot {\bf R}_{mn}} b_{mn}^{00} \cdot t_{0,m}^{\dagger}({\bf k}) t_{0,n}^{\dagger}(-{\bf k}) + h.c.,
\end{eqnarray}
where
\begin{eqnarray}
&&
\epsilon_{PQ}
\nonumber\\
&=&
\sum_{(m,n)}
A_{mn}
\left[
-P_{mn}^{++*} \cdot P_{mn}^{00}
-P_{mn}^{00*} \cdot P_{mn}^{--}
\right.
\nonumber\\
&&
~~~~~~~~~~~~~~
\left.
+Q_{mn}^{+-*} \cdot Q_{mn}^{00}
+Q_{mn}^{00*} \cdot Q_{mn}^{-+}
\right]
+
h.c.
\nonumber\\
&+&
\sum_{(m,n)}
B_{mn}
\left[
-\left| P_{mn}^{++} \right|^2
-\left| P_{mn}^{--} \right|^2
+\left| Q_{mn}^{+-} \right|^2
+\left| Q_{mn}^{-+} \right|^2
\right],
\nonumber\\
\label{eq:epsilon_PQ}
\end{eqnarray}
and
\begin{subequations}
\begin{eqnarray}
&&
a_{mn}^{++}
=
A_{mn} P_{mn}^{00}
+
B_{mn} P_{mn}^{++},
\\
&&
a_{mn}^{--}
=
A_{mn}^* P_{mn}^{00}
+
B_{mn} P_{mn}^{--},
\\
&&
a_{mn}^{00}
=
A_{mn} P_{mn}^{--}
+
A_{mn}^* P_{mn}^{++},
\\
&&
b_{mn}^{+-}
=
-A_{mn} Q_{mn}^{00}
-B_{mn} Q_{mn}^{+-},
\\
&&
b_{mn}^{-+}
=
-A_{mn}^* Q_{mn}^{00}
-B_{mn} Q_{mn}^{-+},
\\
&&
b_{mn}^{00}
=
-A_{mn} Q_{mn}^{-+}
-A_{mn}^* Q_{mn}^{+-}.
\end{eqnarray}
\end{subequations}

A few comments on the mean-field decoupling are in order.
First, it must be noted that there is another possibility in decoupling the last terms with $B_{mn}$ in (\ref{eq:H_quartic}).
It is to decouple them by using the direct channel;
\begin{eqnarray}
 &&
 \left[ t_{+,i}^{\dagger} t_{+,i} - t_{-,i}^{\dagger} t_{-,i} \right]
 \left[ t_{+,j}^{\dagger} t_{+,j} - t_{-,j}^{\dagger} t_{-,j} \right]
 \nonumber\\
 &\rightarrow&
 \left[ \langle t_{+,i}^{\dagger} t_{+,i} \rangle - \langle t_{-,i}^{\dagger} t_{-,i} \rangle \right]
 \left[ t_{+,j}^{\dagger} t_{+,j} - t_{-,j}^{\dagger} t_{-,j} \right]
 \nonumber\\
 &+&
 \left[ t_{+,i}^{\dagger} t_{+,i} - t_{-,i}^{\dagger} t_{-,i} \right]
 \left[ \langle t_{+,j}^{\dagger} t_{+,j} \rangle - \langle t_{-,j}^{\dagger} t_{-,j} \rangle \right]
 \nonumber\\
 &-&
 \left[ \langle t_{+,i}^{\dagger} t_{+,i} \rangle - \langle t_{-,i}^{\dagger} t_{-,i} \rangle \right]
 \left[ \langle t_{+,j}^{\dagger} t_{+,j} \rangle - \langle t_{-,j}^{\dagger} t_{-,j} \rangle \right],
 \nonumber\\
\end{eqnarray}
where $i$ and $j$ mean $({\bf r},m)$ and $({\bf r}+{\bf R}_{mn},n)$, respectively.
However, the direct channel mean-fields do not make any effect because they cancel each other so that their contribution becomes zero.
The cancellation originates from the time reversal symmetry of the original Hamiltonian (\ref{eq:H_J_H_DM}).
Due to the symmetry, $\langle t_{+,i}^{\dagger} t_{+,i} \rangle = \langle t_{-,i}^{\dagger} t_{-,i} \rangle$ and the direct channel mean-fields have no effect.
Secondly, the above mean-field decoupling is intended for the Hamiltonian with nonzero DM interactions ($d_z \ne 0$).
For the case without the DM interactions ($d_z=0$), reader is referred to Ref. 32.



\end{document}